\documentclass[twocolumn]{aastex631}
\usepackage{CJK}
\usepackage{xcolor}
\usepackage{amsmath}
\usepackage{longtable}
\usepackage{upgreek}
\usepackage{xspace}

\usepackage[utf8]{inputenc}

\shorttitle{PRSs in Dwarf Galaxies}
\shortauthors{Dong et al.}

\def\ra#1#2#3{#1$^{\rm h}$#2$^{\rm m}$#3$^{\rm s}$}
\def\dec#1#2#3{$#1^\circ#2'#3''$}
\newcommand{\dmunit}{pc~cm$^{-3}$\xspace}
\newcommand{\nsamp}{9}

\begin{document}
\begin{CJK*}{UTF8}{gbsn}

\title{A Radio Study of Persistent Radio Sources in Nearby Dwarf Galaxies: \\ Implications for Fast Radio Bursts}

\newcommand{\NU}{\affiliation{Center for Interdisciplinary Exploration and Research in Astrophysics (CIERA) and Department of Physics and Astronomy, Northwestern University, Evanston, IL 60208, USA}}

\newcommand{\CSIRO}{\affiliation{CSIRO, Space and Astronomy, PO Box 76, Epping, NSW 1710, Australia}}

\newcommand{\ASTRON}{\affiliation{ASTRON, Netherlands Institute for Radio Astronomy, Oude Hoogeveensedijk 4, 7991 PD Dwingeloo, The Netherlands}}

\newcommand{\UAmsterdam}{\affiliation{Anton Pannekoek Institute for Astronomy, University of Amsterdam, Science Park 904, 1098 XH, Amsterdam, The Netherlands}}

\newcommand{\umn}{\affiliation{School of Physics and Astronomy, University of Minnesota, Minneapolis, Minnesota 55455, USA}}

\newcommand{\CfA}{\affiliation{Center for Astrophysics\:$|$\:Harvard \& Smithsonian, 60 Garden St. Cambridge, MA 02138, USA}}

\newcommand{\CCA}{\affiliation{Center for Computational Astrophysics, Flatiron Institute, 162 W. 5th Avenue, New York, NY 10011, USA}}

\newcommand{\ColumbiaPhysics}{\affiliation{Department of Physics and Columbia Astrophysics Laboratory, Columbia University, New York, NY 10027, USA}}

\newcommand{\ColumbiaAstro}{\affiliation{Department of Astronomy and Columbia Astrophysics Laboratory, Columbia University, New York, NY 10027, USA}}

\newcommand{\Cahill}{\affiliation{Cahill Center for Astronomy and Astrophysics, California Institute of Technology, Pasadena, CA 91106, USA}}

\newcommand{\Montana}{\affiliation{eXtreme Gravity Institute, Department of Physics, Montana State University, Bozeman, MT 59717, USA}}

\newcommand{\Boulder}{\affiliation{Center for Astrophysics and Space Astronomy, Department of Astrophysical and Planetary Sciences, University of Colorado, 389 UCB, Boulder, CO 80309-0389, USA}}

\newcommand{\Princeton}{\affiliation{Department of Astrophysical Sciences, Princeton University, Princeton, NJ 08544, USA}}

\newcommand{\THEA}{\affiliation{Theoretical High Energy Astrophysics (THEA) Group, Columbia University, New York, New York 10027, USA}}

\newcommand{\MIT}{\affiliation{MIT Kavli Institute for Astrophysics and Space Research, Massachusetts Institute of Technology, Cambridge, MA 02139}}

\newcommand{\JIVE}{\affiliation{Joint Institute for VLBI ERIC, Oude Hoogeveensedijk 4, 7991~PD, Dwingeloo, The Netherlands}}

\author[0000-0002-9363-8606]{Y.~Dong (董雨欣)}
\NU

\author[0000-0003-0307-9984]{T.~Eftekhari}
\altaffiliation{NHFP Einstein Fellow}
\NU

\author[0000-0002-7374-935X]{W.~Fong}
\NU

\author[0000-0003-3460-506X]{S.~Bhandari}
\ASTRON
\JIVE
\UAmsterdam
\CSIRO

\author[0000-0002-9392-9681]{E.~Berger} 
\CfA

\author[0000-0001-9381-8466]{O.S.~Ould-Boukattine}
\ASTRON
\UAmsterdam

\author[0000-0003-2317-1446]{J.W.T.~Hessels}
\ASTRON
\UAmsterdam

\author[0000-0002-5519-9550]{N.~Sridhar}
\ColumbiaAstro
\THEA
\Cahill

\author[0000-0001-7158-614X]{A.~Reines}
\Montana

\author[0000-0001-8405-2649]{B.~Margalit}
\umn

\author[0000-0003-2511-2060]{J.~Darling}
\Boulder

\author[0000-0002-5025-4645]{A.~C.~Gordon}
\NU

\author{J.E.~Greene}
\Princeton

\author[0000-0002-5740-7747]{C.~D.~Kilpatrick}
\NU

\author[0000-0001-9814-2354]{B.~Marcote} 
\JIVE

\author[0000-0002-4670-7509]{B.~D.~Metzger}
\ColumbiaPhysics
\THEA
\CCA

\author[0000-0003-0510-0740]{K.~Nimmo}
\MIT

\author[0000-0002-2028-9329]{A.~E.~Nugent}
\NU

\author[0000-0002-5195-335X]{Z.~Paragi}
\JIVE

\author[0000-0003-3734-3587]{P.K.G.~Williams}
\CfA

\begin{abstract}
We present 1 -- 12\,GHz Karl G. Jansky Very Large Array observations of \nsamp~off-nuclear persistent radio sources (PRSs) in nearby ($z \lesssim 0.055$) dwarf galaxies, along with high-resolution European very-long baseline interferometry (VLBI) Network (EVN) observations for one of them at 1.7\,GHz. We explore the plausibility that these PRSs are associated with fast radio burst (FRB) sources by examining their properties --- physical sizes, host-normalized offsets, spectral energy distributions (SEDs), radio luminosities, and light curves --- and compare them to those of the PRSs associated with FRBs\,20121102A and 20190520B, two known active galactic nuclei (AGN), and one likely AGN in our sample with comparable data, as well as other radio transients exhibiting characteristics analogous to FRB-PRSs. We identify a single source in our sample, J1136$+$2643, as the most promising FRB-PRS, based on its compact physical size and host-normalized offset. We further identify two sources, J0019$+$1507 and J0909$+$5655, with physical sizes comparable to FRB-PRSs, but which exhibit large offsets and flat spectral indices potentially indicative of a background AGN origin. We test the viability of neutron star wind nebula and hypernebula models for J1136$+$2643 and find that the physical size, luminosity, and SED of J1136$+$2643 are broadly consistent with these models. Finally, we discuss the alternative interpretation that the radio sources are instead powered by accreting massive black holes, and we outline future prospects and follow-up observations for differentiating between these scenarios.
\end{abstract}

\section{Introduction}\label{sec:intro}
The extragalactic radio sky is primarily composed of galaxies powered by recent star formation and active galactic nuclei (AGN; \citealt{Condon12}). At low flux densities, the abundance of faint radio sources is dominated by ongoing star formation in spiral and irregular galaxies \citep{JJCondon92, Padovani16}, while radio-emitting AGN powering relativistic jets in elliptical galaxies become prominent at higher flux densities \citep{Condon12, Magliocchetti22}. However, recent investigations into extragalactic radio sources over the past decade have unveiled a distinct, third population of compact, persistent radio sources (PRSs) that preferentially reside in dwarf host galaxy environments. 

First coined in association with fast radio bursts (FRBs; \citealt{Law22}), the PRS population is variegated into two sub-classes: one related to intermediate mass (100  M$_{\odot}$ -- 10$^{5} M_{\odot}$) black holes (IMBHs; \citealt{Maccarone04,Greene20}) and the other linked to FRBs and other transient phenomena. In the context of IMBHs, numerous studies have relied on multiwavelength accretion signatures to gain demographic statistics on IMBHs, leading to the identification of hundreds of AGN candidates in dwarf galaxies \citep{Barth08, Reines13, Moran14, Sartori15,Baldassare20}, with some appearing as off-nuclear wanderers \citep{Mezcua20,Reines20}. Among those AGN candidates identified in dwarf galaxies, some exhibit radio emission and appear as compact radio sources. However, it remains challenging to firmly establish the presence of such IMBHs, as they cannot be easily distinguished from background interlopers and thus require follow-up observations at other wavelengths.

At the same time, a subset of the PRS population has been associated with transient phenomena. Indeed, two luminous compact PRSs have been unambiguously associated with FRBs in dwarf galaxies: FRBs\,20121102A and 20190520B \citep{Chatterjee17, Niu22}. FRBs are luminous, millisecond-duration pulses of coherent radio emission whose origin is still broadly debated despite hundreds of sources identified thus far \citep{Lorimer07, Bannister17, Law18REALFAST, CHIME_catalog, Petroff22, Zhang23Review}. The presence of PRSs associated with two FRBs have provided some of the most stringent constraints on their progenitors to date. In particular, their possible preference in dwarf environments is consistent with that of long $\gamma$-ray bursts (LGRBs) and superluminous supernovae (SLSNe) resulting from the explosion of young massive stars \citep{Murase16, Metzger17}. In both cases, the radio counterparts have been interpreted as emission from nebulae surrounding a central engine such as a magnetar \citep{Beloborodov17, Kashiyama17, Margalit18, Li20} or an accreting compact object \citep{Sridhar+22, Sridhar+24a}. As demonstrated by the variable Faraday rotation measure (RM) of the bursts, they also reside in turbulent magneto-ionic environments \citep{Michilli18, Anna-Thomas23}.

Following the identification of the FRB-PRS class, a population of slowly evolving radio sources in dwarf galaxies were discovered, including the decades-long radio transient FIRST J141918.9$+$394036 (hereafter J1419$+$3940) which has been postulated as an orphan LGRB afterglow \citep{Law18,Marcote19}; the SLSN PTF10hgi \citep{Eftekhari19}, whose PRS was identified nearly a decade post-explosion; and the pulsar wind nebula (PWN) candidate J113706.19$-$033737.3 (hereafter VT 1137$-$0337; \citealt{DDong23}). Together, these discoveries exemplify the rich diversity of extragalactic radio sources within dwarf galaxies.

PRS discovery methods have thus far been largely heterogeneous, ranging from targeted searches for radio emission from the locations of known transients \citep{Ofek17, Eftekhari19, Law19PTF} to systematic searches in nearby dwarf galaxies \citep{Reines20,Vohl23}. Recently, \cite{Reines20} presented a sample of 13 PRSs discovered in dwarf galaxies, some of which are offset from their galactic centers. They rule out star-forming H\,II regions, supernova remnants (SNRs), and radio supernovae (SNe) as the origin of the radio emission, and instead conclude that these PRSs represent a population of off-nuclear accreting IMBHs. Conversely, \cite{Eftekhari20} argued for an alternative interpretation in which these sources are instead analogs to PRSs associated with FRBs, as evidenced by the shared similarities in the observed radio properties with the PRS coincident with FRB\,20121102A (i.e., PRS\,20121102A).
Studies like \citet{Reines20} take advantage of the shared dwarf host galaxy environments of PRSs as a criterion for systematic PRS discovery, as opposed to relying on serendipitous transient discovery. Here, we further explore the possibility that these radio sources share the same origin as FRB-PRSs using multi-frequency radio observations obtained with the Karl G. Jansky Very Large Array (VLA) and the European very-long baseline interferometry (VLBI) Network (EVN). 

The paper is organized as follows. We present our sample of radio sources in Section \ref{sec:sample} and report a compilation of VLA, EVN, and archival radio observations in Section \ref{sec:obs}. In Section \ref{sec:res}, we present the properties of the sources, including their physical sizes, host-normalized offsets, spectral energy distributions (SEDs), and light curves, and compare them to those of PRSs\,20121102A and 20190520B along with radio transients J1419$+$3940, PTF10ghi, and VT 1137$-$0337. Based on these properties, we identify a single source, J1136$+$2643, as the most compelling FRB-PRS candidate. In Section \ref{sec:models}, we test the viability of our FRB-PRS candidate in the context of FRB progenitor models. Finally, in Section \ref{sec:AGNorigin}, we consider the alternative interpretation that the radio sources are instead powered by AGN, and summarize our conclusions in Section \ref{sec:conclusions}. Throughout the paper, we adopt the \textit{Planck} cosmological parameters for a flat $\Lambda$CDM universe, with $H_{0}$ = 67.66 km s$^{-1}$ Mpc$^{-1}$, $\Omega_m = 0.310$, and $\Omega_{\lambda} = 0.690$ \citep{Planck18} for luminosity and physical size calculations.

\begin{deluxetable*}{cccccccc}[t!]
\linespread{1.2}
\tabletypesize{\footnotesize}
\tablecaption{VLA and EVN Observations of the Persistent Radio Sources}
\tablecolumns{6}
\tablewidth{0pt}
\label{tab:vla_obs}
\tablehead{
\colhead{Source} &
\colhead{ID$^\dagger$} &
\colhead{Telescope} & 
\colhead{Date} &
\colhead{Flux Calibrator} &
\colhead{Phase Calibrator} &
\colhead{Fringefinder} & 
\colhead{t$_{\mathrm{int}}$} \\
\colhead{} &
\colhead{} & 
\colhead{} & 
\colhead{(mm/dd/yyyy)} & 
\colhead{} & 
\colhead{} &  
\colhead{} & 
\colhead{(minutes)}
}
\startdata
J0019$+$1507 & 2 & VLA-BnA & 11/24/2020 & 3C48 & J0010+1724 & $\ldots$ & 12.40 \\
& & EVN & 04/13/2021 & $\ldots$ & J1028+0255 & J0237+2848 & 194.6 \\
J0106$+$0046 & 6 & VLA-BnA & 11/24/2020 & 3C48 & J0125$-$0005 & $\ldots$ & 12.47 \\
J0903$+$4824 & 25 & VLA-BnA & 11/25/2020 & 3C147 & J0920$+$4441 & $\ldots$ & 13.07 \\
J0909$+$5655 & 28 & VLA-BnA & 11/25/2020 & 3C147 & J0921+6215 & $\ldots$ & 12.47 \\
J0931$+$5633 & 33 & VLA-BnA & 11/25/2020 & 3C147 & J0921+6215 & $\ldots$ & 12.40 \\
J1027$+$0112 & 48 & VLA-BnA & 11/25/2020 & 3C147 & J1024$-$0052 & $\ldots$ & 11.80 \\
& & EVN & 04/13/2021 & $\ldots$ & J0010+1724 & J0237+2848 & 196.0 \\
J1136$+$1252 & 64 & VLA-BnA & 11/25/2020 & 3C286 & J1120+1420 & $\ldots$ & 13.60 \\
& & EVN & 06/22/2021 & $\ldots$ & J1132+1023 & J1058+0133 & 168.34 \\
J1136$+$2643 & 65 & VLA-BnA & 11/25/2020 & 3C286 & J1125+2610 & $\ldots$ &  10.90 \\
J1200$-$0341 & 77 & VLA-BnA & 11/25/2020 & 3C286 & J1150$-$0023 & $\ldots$ & 12.40 \\
J1220$+$3020 & 82 & VLA-BnA & 11/25/2020 & 3C286 & J1221+2813 & $\ldots$ & 11.53 \\
J1226$+$0815 & 83 & VLA-BnA & 11/25/2020 & 3C286 & J1239+0730 & $\ldots$ & 11.77 \\
J1253$-$0312 & 92 & VLA-BnA & 11/25/2020 & 3C286 & J1246$-$0730 & $\ldots$ & 11.77
\enddata
\tablecomments{$^\dagger$ Galaxy identification number as assigned in \cite{Reines20}.}
\end{deluxetable*}

\section{Sample of Compact Radio Sources}\label{sec:sample}
We derive our sample from a catalogue of compact radio sources initially discovered by \cite{Reines20} using the VLA. In particular, they identified a sample of 19 compact radio sources that exhibited the strongest evidence for accreting wandering IMBHs while also being inconsistent with a star formation origin. We aim to leverage the similarities between the known PRSs and the compact radio sources found in dwarf galaxies, some of which demonstrate spatial offsets from their host centers akin to the PRSs. Motivated by the dwarf host environment of PRSs, we only consider 13 of the initial 19 that were in \textit{bona fide} (M$_*$ $\lesssim$ 3 $\times$ 10$^9$ M$_\odot$) dwarf galaxies with robust redshifts (``Sample A'' in \citealt{Reines20}). However, as discussed in \cite{Reines20} and \cite{Eftekhari20}, only one out of the 13 dwarf galaxies, J0906+5610, is securely in the AGN region on the Baldwin-Phillips-Terlevich (BPT) diagram \citep{Baldwin81} and was previously identified as a broad-line AGN \citep{Reines13}. We therefore exclude J0906$+$5610 as a potential FRB-PRS candidate in this work. The rest of the sample falls in the region occupied by star-forming galaxies.

More recently, among the parent sample of 13 radio sources in dwarf galaxies, J1220$+$3020 and J1136$+$1252 have been confirmed as AGN based on spectroscopic data. In particular, J1220$+$3020 shows the presence of coronal line [Fe\,X] and enhanced [O\,I] emission \citep{Molina21}, while J1136$+$1252 is associated with an optical counterpart in deep \textit{Hubble Space Telescope} (\textit{HST}) imaging at a spectroscopic redshift of $z = 0.76$ (Sturm et al. 2024, \textit{in prep.}), pointing to a background quasar origin. Nevertheless, we include them in our analysis as they provide useful comparisons by representing the known AGN in our sample. Lastly, archival optical imaging reveals a red optical counterpart at the position of J1027$+$0112, hinting at a possible background AGN origin (Section~\ref{sec:optical_image}). 

Our final sample therefore includes \nsamp~radio sources excluding the two confirmed AGN and one likely AGN. In the following sections, we refer to all \nsamp~sources as PRSs and explore them as potential FRB-PRSs. To constrain the size of the sources in our sample down to sub-arcsecond scales, we separately construct a subset of three targets for follow-up with the EVN. We expand upon the details of this sub-sample in Section~\ref{sec:VLBI_obs}.

\section{Observations} \label{sec:obs}
\subsection{VLA Continuum Observations}\label{sec:VLA_obs}
We observed our sample of 12 radio sources with the VLA under program 20B-228 (PI: T.~Eftekhari) in the hybrid BnA configuration. The sample was grouped into three scheduling blocks ranging between $\sim$ 1--2 hours each, with on-source integration times of $\approx$ 11--14 min per source. The observations were taken between 2020 November 24 and 25 UTC in four frequency bands centered at 1.5 (L-band), 3.0 (S-band), 6.0 (C-band), and 10.0~GHz (X-band). At the higher frequencies (4$-$12 GHz), we utilized the 3-bit samplers which provide the full 4~GHz of bandwidth across the observing band. At the lower frequencies, we employed the 8-bit samplers to achieve 1 and 2~GHz of bandwidth at L- and S-band, respectively.

We processed the data using the standard VLA pipeline (version 2022.2.0.64) as part of the Common Astronomy Software Applications (CASA; \citealt{casa, Bemmel_2022}) software package. We performed bandpass and flux density calibration using 3C48, 3C147, and 3C286 in each scheduling block, respectively. Details of the observations, including complex gain calibrators for individual sources and on-source integration times are listed in Table~\ref{tab:vla_obs}. We split the calibrated data in each frequency band into two subbands and imaged each subband separately using CASA's \texttt{tclean} task out to the first null of the primary beam. We set a pixel scale of 0.43, 0.22, 0.11, and 0.07\arcsec pixel$^{-1}$ at 1.5, 3, 6, and  10\,GHz, respectively. We performed deconvolution using standard gridders, a Briggs visibility weighting scheme with a robust parameter of 0, and multi-term multi-frequency synthesis (MTMFS; \citealt{RC11}) with two Taylor terms.

We extracted the flux densities at the source positions and image root-mean-square (rms) values using the \texttt{imtool} task as part of the \texttt{pwkit} package \citep{pwkit}. We also applied an additional 5$\%$ error to the measured uncertainties to account for the accuracy of the absolute flux calibration scale of the VLA \citep{PB17}. We detect unresolved radio emission at the positions of all the sources in our sample with the exception of J0903+4824, which appears to be resolved into two components. A summary of our VLA observations is given in Table~\ref{tab:vla_data}.

\begin{figure}[t!]
    \centering
    \includegraphics[width=0.45\textwidth,clip,trim={0cm 0cm 0cm 1cm}]{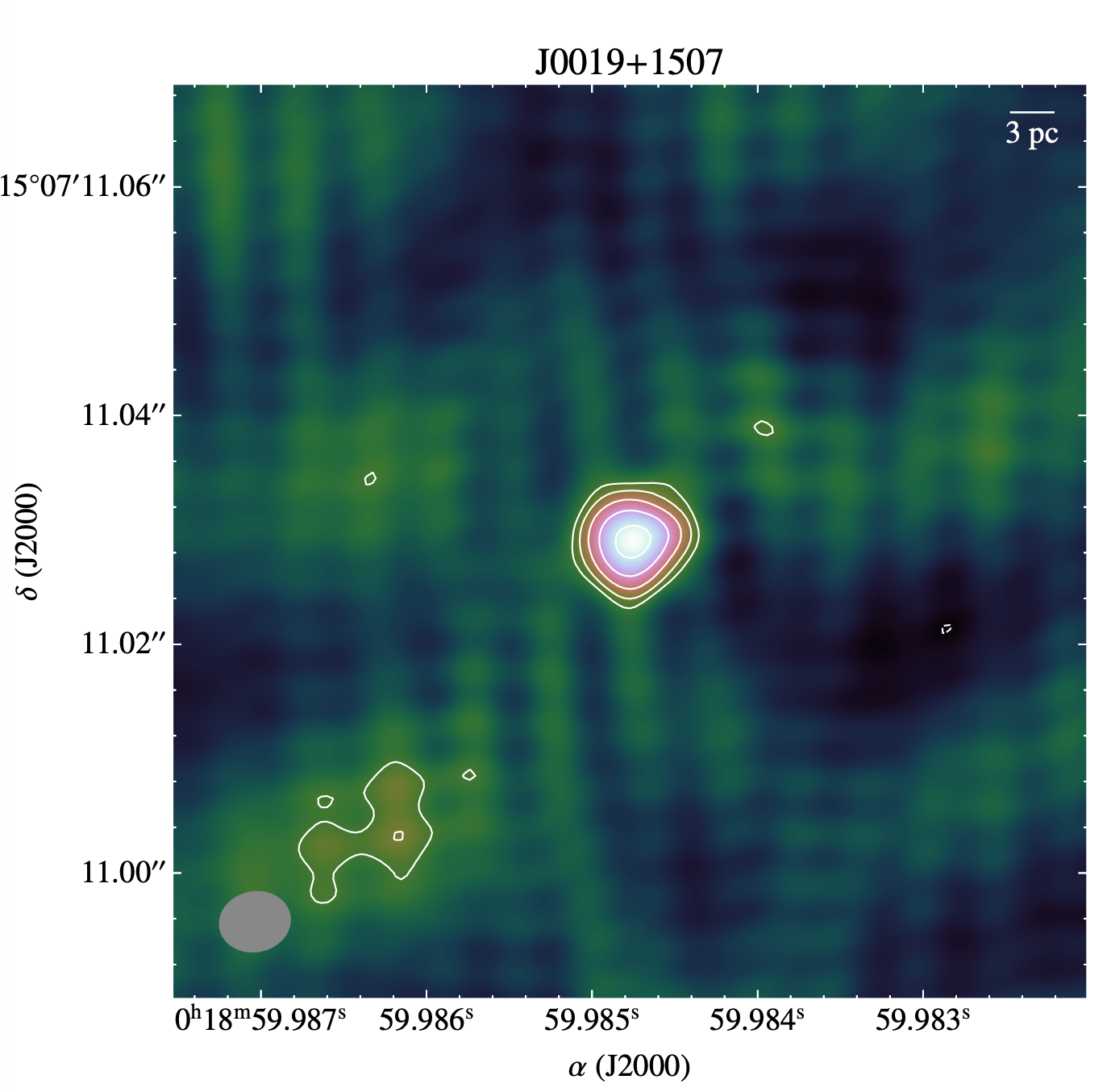}
    \caption{EVN-detected source J0019$+$1507 at 1.7\,GHz with contour lines depicting rms levels starting from $3.5\sigma$, where $\sigma$ is 50\,$\upmu$Jy\,beam$^{-1}$. A small bar in the top-right of the image shows the physical scale assuming that the sources are at the same redshift of the associated dwarf galaxy. The synthesized beam of size {6\,mas $\times$ 5\,mas} is represented by the grey ellipse in the lower left corner.}
    \label{fig:EVN_sources}
\end{figure}

\subsection{EVN VLBI Observations} \label{sec:VLBI_obs}

We observed three sources from our sample with the EVN, including e-MERLIN stations. Sources J0019$+$1507 and J1027$+$0112 were observed on 2021 April 13 UTC (EO018A; PI: O.~Ould-Boukattine), and source J1136$+$1252 was observed on 2021 June 22 UTC (EO018B; PI: O.~Ould-Boukattine). The sample was chosen to be unresolved by the VLA, have a minimum flux density of 2.5\,mJy at 9\,GHz and a relatively flat spectral index. In addition, they exhibit relatively small physical offset from the center of their host galaxies. The observations included regular EVN stations.\footnote{Full list of EVN stations: Mark II Jodrell Bank (Jb), Westerbork single dish (Wb, RT1), Effelsberg (Ef), Medicina (Mc),  Onsala (O8), Tianma (T6), Toru\'n (Tr), Hartebeesthoek (Hf), Svetloe (Sv), Zelenchukskaya (Zc), Badary (Bd), Irbene (Ir), Sardinia (Sr) and e-MERLIN stations: Cambridge (Cm), Darnhall (Da), Knockin (Kn) and Pickmere (Pi). Of these, the Russian stations (Sv, Zc and Bd), Sr and Cm did not participate in the second session.} At the time of observation, the likely background AGN J1027+0112 and the confirmed background AGN J1136$+$1252 were still considered FRB-PRS candidates with unknown origins. All of our observations were conducted at 1.7\,GHz (L-band) with $4\times 32$\,MHz sub-bands. Details of the observations, including fringe finder and phase calibrator sources as well as on-source integration times are listed in Table~\ref{tab:vla_obs}.

We correlated the interferometric data using the software correlator SFXC \citep{Keimpema+15} at the Joint Institute for VLBI ERIC (JIVE), with an integration time of 2s and 128 channels per 32 MHz subband. We calibrated the data following standard procedures in AIPS \citep{AIPS} and CASA \citep{casa,Bemmel_2022}. Using the \texttt{FITLD} task, we loaded the correlated visibilities in FITS-IDI format into AIPS and applied the \textit{a-priori} amplitude calibration and \textit{a-priori} flagging table from the table generated by the EVN AIPS pipeline and used \texttt{VLBATECR} in AIPS to correct for ionospheric dispersive delays. We then performed the rest of the calibration in CASA following the steps described in \citet{Bhandari+23b}.

We imaged all three targets using {\sc Difmap} \citep{DIFMAP}. Source J0019$+$1507 along with the background sources, J1027$+$0112 and J1136$+$1252, were all detected in our EVN observations. Figure~\ref{fig:EVN_sources} shows the continuum image of J0019$+$1507 at 1.7\,GHz (see the \hyperref[sec:appendix]{Appendix} for EVN images of sources J1027$+$0112 and J1136$+$1252). We constrained the apparent angular sizes and measured the flux densities of all three sources in {\sc Difmap} by $\chi^2$--fitting of a circular Gaussian model in the $uv$-plane using \texttt{modelfit}. The flux densities of the sources are presented in Table\,\ref{tab:vla_data}. We note that the uncertainty on the flux density is the quadrature sum of the rms noise and 15\% of the absolute flux error, which is typical for VLBI observations.

\startlongtable
\setlength{\LTleft}{-5cm}
\begin{deluxetable*}{l|cccccccc}
\tabletypesize{\footnotesize}
\label{tab:vla_data}
\tablecolumns{9}
\tablewidth{0pc}
\tablecaption{Radio Observation Catalog
\label{tab:prospectres}}
\tablehead{
\colhead{Source}	 &
\colhead{R.A.} &
\colhead{Decl.} &
\colhead{Frequency} &
\colhead{Beam Size} &
\colhead{Beam Angle} &
\colhead{RMS} &
\colhead{Flux Density} \\
\colhead{} &
\colhead{(J2000)} & 
\colhead{(J2000)} &
\colhead{(GHz)} &
\colhead{(arcsec$^{2}$)} &
\colhead{(deg)} & 
\colhead{($\upmu$Jy/beam)}&
\colhead{(mJy)}
}
\startdata
J0019$+$1507 & \ra{00}{18}{59.985} & \dec{+15}{07}{11.02} & 1.3 & 4.04 $\times$ 1.24 & $-$80.36 & 42 & 2.19 $\pm$ 0.13 \\ % COORDS from Sargent+
& & & 1.7 & 0.006$\times$0.005 & $-$78.85 & 50 & 1.30 $\pm$ 0.20 \\
& & & 1.8 & 2.73 $\times$ 0.93 & $-$82.03 & 54 & 2.00 $\pm$ 0.13 \\
& & & 2.6 & 2.13 $\times$ 0.62 & $-$80.32 & 40 & 2.24 $\pm$ 0.13 \\
& & & 3.4 & 1.78 $\times$ 0.46 & $-$79.50 & 35 & 2.23 $\pm$ 0.12 \\
& & & 5 & 1.08 $\times$ 0.32 & $-$81.05 & 38 & 2.08 $\pm$ 0.12 \\
& & & 7 & 0.78 $\times$ 0.22 & $-$81.38 & 40 & 2.13 $\pm$ 0.12 \\
& & & 9 & 0.70 $\times$ 0.16 & $-$81.87 & 38 & 2.06 $\pm$ 0.11 \\
& & & 11 & 0.52 $\times$ 0.14 & $-$82.65 & 30 & 2.00 $\pm$ 0.11  \\
\hline
J0106$+$0046 & \ra{01}{06}{07.308} & \dec{+00}{46}{34.320} & 1.3 & 5.97 $\times$ 1.18 & $-$69.56 & 66 & 1.91 $\pm$ 0.13 \\
& & & 1.8 & 4.37 $\times$ 0.90 & $-$71.54 & 66 & 1.48 $\pm$ 0.12 \\
& & & 2.6 & 3.00 $\times$ 0.65 & $-$70.63 & 38 & 0.96 $\pm$ 0.07 \\
& & & 3.4 & 2.13 $\times$ 0.52 & $-$70.45 & 37 & 0.85 $\pm$ 0.07 \\
& & & 5 & 1.48 $\times$ 0.34 & $-$71.01 & 37 & 0.59 $\pm$ 0.06 \\
& & & 7 & 1.14 $\times$ 0.22 & $-$74.21 & 21 & 0.45 $\pm$ 0.04 \\
& & & 9 & 0.86 $\times$ 0.19 & $-$73.98 & 24 & 0.28 $\pm$ 0.04 \\
& & & 11 & 0.75 $\times$ 0.14 & $-$74.92 & 30 & 0.24 $\pm$ 0.04 \\
\hline
J0903$+$4824 & \ra{09}{03}{12.967} & \dec{+48}{24}{13.716} & 1.3 & 1.94 $\times$ 1.16 & $-$65.49 & 377 & 6.51 $\pm$ 0.62 \\
& & & 1.8 & 1.40 $\times$ 0.84 & $-$66.80 & 415 & 4.18 $\pm$ 0.62 \\
& & & 2.6 & 0.98 $\times$ 0.60 & $-$66.00 & 245 & 2.48 $\pm$ 0.37 \\
& & & 3.4 & 0.75 $\times$ 0.46 & $-$65.96 & 148 & 1.79 $\pm$ 0.23 \\
& & & 5 & 0.52 $\times$ 0.32 & $-$60.08 & 79 & 1.46 $\pm$ 0.16 \\
& & & 7 & 0.53 $\times$ 0.23 & 87.09 & 19 & 0.99 $\pm$ 0.06  \\
& & & 9 & 0.29 $\times$ 0.17 & $-$68.23 & 39 & 0.76 $\pm$ 0.09 \\
& & & 11 & 0.23 $\times$ 0.14 & $-$69.41 & 19 & 0.68 $\pm$ 0.06 \\
\hline
J0909$+$5655 & \ra{09}{09}{08.689} & \dec{+56}{55}{19.750} & 1.3 & 2.00 $\times$ 1.19 & $-$60.16 & 56 & 1.53 $\pm$ 0.11\\ % Coords from Sargent+
& & & 1.8 & 1.43 $\times$ 0.87 & $-$60.35 & 58 & 1.37 $\pm$ 0.11 \\
& & & 2.6 & 1.00 $\times$ 0.62 & $-$65.65 & 31 & 1.32 $\pm$ 0.08 \\
& & & 3.4 & 0.76 $\times$ 0.48 & $-$64.39 & 33 & 1.29 $\pm$ 0.08 \\
& & & 5 & 0.54 $\times$ 0.33 & $-$64.79 & 34 & 1.09 $\pm$ 0.07 \\
& & & 7 & 0.74 $\times$ 0.23 & 76.50 & 19 & 0.91 $\pm$ 0.05 \\
& & & 9 & 0.30 $\times$ 0.17 & $-$65.11 & 19 & 0.73 $\pm$ 0.05 \\
& & & 11 & 0.24 $\times$ 0.15 &$-$71.81 & 22 & 0.65 $\pm$ 0.04 \\ 
\hline
J0931$+$5633 & \ra{09}{31}{38.419} & \dec{+56}{33}{19.872} & 1.3 & 2.01 $\times$ 1.20 & $-$60.15 & 65 & 9.37 $\pm$ 0.48 \\
& & & 1.8 & 1.44 $\times$ 0.87 & $-$60.22 & 54 & 7.70 $\pm$ 0.39 \\
& & & 2.6 & 1.05 $\times$ 0.59 & $-$59.62 & 51 & 5.84 $\pm$ 0.30 \\
& & & 3.4 & 0.80 $\times$ 0.46 & $-$58.73 & 40 & 4.90 $\pm$ 0.25 \\
& & & 5 & 0.53 $\times$ 0.33  & $-$64.34 & 97 & 3.63 $\pm$ 0.23 \\
& & & 7 & 0.74 $\times$ 0.23 & 77.24 & 31 & 2.75 $\pm$ 0.14 \\
& & & 9 & 0.29 $\times$ 0.17 & $-$64.94 & 27 & 2.17 $\pm$ 0.11 \\
& & & 11 & 0.24 $\times$ 0.15 & $-$71.22 & 21 & 1.88 $\pm$ 0.10 \\ 
\hline
J1027$+$0112 & \ra{10}{27}{41.378} & \dec{+01}{12}{06.444} & 1.3 & 3.75 $\times$ 1.40 & 81.13 & 119 & 4.38 $\pm$ 0.28 \\
& & & 1.7 & 0.006$\times$0.004 & 21.93 & 62 & 2.80 $\pm$ 0.40 \\
& & & 1.8 & 2.9 $\times$ 0.95 & 84.12 & 70 & 3.94 $\pm$ 0.22 \\
& & & 2.6 & 2.08 $\times$ 0.69 & 84.13 & 65 & 3.23 $\pm$ 0.19 \\
& & & 3.4 & 1.77 $\times$ 0.54 & 82.66 & 43 & 3.46 $\pm$ 0.18 \\
& & & 5 & 1.03 $\times$ 0.38 & 74.97 & 29 & 3.16 $\pm$ 0.16 \\
& & & 7 & 0.98 $\times$ 0.24 & 80.25 & 22 & 3.07 $\pm$ 0.16 \\
& & & 9 & 0.57 $\times$ 0.20 & 78.41 & 25 & 2.88 $\pm$ 0.15 \\
& & & 11 & 0.50 $\times$ 0.15 & 82.75 & 28 & 2.72 $\pm$ 0.14 \\
\hline
J1136$+$1252 & \ra{11}{36}{48.526} & \dec{+12}{52}{39.900} & 1.3 & 8.15 $\times$ 1.27 & 58.25 & 77 & 2.78 $\pm$ 0.18 \\
& & & 1.7 & 0.022 $\times$ 0.005 & 80.10 & 65 & 1.60 $\pm$ 0.30 \\ 
& & & 1.8 & 5.58 $\times$ 0.82 & 59.08 & 76 &  2.48 $\pm$ 0.16 \\
& & & 2.6 & 3.74 $\times$ 0.71 & 57.37 & 51 & 2.48 $\pm$ 0.14 \\
& & & 3.4 & 2.68 $\times$ 0.47 & 59.50 & 49 & 2.62 $\pm$ 0.14 \\
& & & 5 & 2.22 $\times$ 0.34 & 59.02 & 34 & 3.00 $\pm$ 0.16 \\
& & & 7 & 1.56 $\times$ 0.25 & 57.91 & 41 & 3.16 $\pm$ 0.17 \\
& & & 9 & 0.56 $\times$ 0.23 & $-$88.97 & 103 & 3.12 $\pm$ 0.21 \\
& & & 11 & 0.46 $\times$ 0.18 & $-$89.03 & 121 & 3.19 $\pm$ 0.23 \\
\hline
J1136$+$2643 & \ra{11}{36}{42.576} & \dec{+26}{43}{35.652} & 1.3 & 7.19 $\times$ 1.35 & 55.80 & 79 & 1.33 $\pm$ 0.13 \\
& & & 1.8 & 4.77 $\times$ 0.87 & 57.30 & 67 & 1.64 $\pm$ 0.13 \\
& & & 2.6 & 3.44 $\times$ 0.70 & 56.78 & 43 & 1.99 $\pm$ 0.12 \\
& & & 3.4 & 2.36 $\times$ 0.49 & 57.25 & 36 & 1.88 $\pm$ 0.11 \\
& & & 5 & 0.87 $\times$ 0.32 & 74.83 & 26 & 1.48 $\pm$ 0.08 \\
& & & 7 & 1.38 $\times$ 0.24 & 56.56 & 27 & 0.95 $\pm$ 0.06 \\
& & & 9 & 1.14 $\times$ 0.20 & 55.92 & 22 & 0.63 $\pm$ 0.04 \\
& & & 11 & 0.97 $\times$ 0.15 & 56.43 & 26 & 0.48 $\pm$ 0.04 \\ 
\hline
J1200$-$0341 & \ra{12}{00}{58.301} & \dec{-03}{41}{18.456} & 1.3 & 4.25 $\times$ 1.44 & 64.98 & 82 & 4.67 $\pm$ 0.26 \\
& & & 1.8 & 7.61 $\times$ 0.94 & 55.92 & 91 & 3.93 $\pm$ 0.23 \\
& & & 2.6 & 5.42 $\times$ 0.68 & 56.10 & 49 & 3.41 $\pm$ 0.18 \\
& & & 3.4 & 3.86 $\times$ 0.49 & 56.10 & 38 & 2.92 $\pm$ 0.16 \\
& & & 5 & 3.06 $\times$ 0.35 & 55.98 & 31 & 2.25 $\pm$ 0.12 \\
& & & 7 & 0.79 $\times$ 0.25 & 66.19 & 34 & 1.84 $\pm$ 0.10 \\
& & & 9 & 2.06 $\times$ 0.20 & 55.12 & 30 & 1.65 $\pm$ 0.09 \\
& & & 11 & 1.67 $\times$ 0.16 & 56.25 & 30 & 1.43 $\pm$ 0.08 \\ 
\hline
J1220$+$3020 & \ra{12}{20}{11.266} & \dec{+30}{20}{08.304} & 1.3 & 6.56 $\times$ 1.39 & 56.45 & 75 & 0.90 $\pm$ 0.11 \\
& & & 1.8 & 4.27 $\times$ 0.90 & 56.97 & 71 & 0.95 $\pm$ 0.11 \\
& & & 2.6 & 3.07 $\times$ 0.69 & 57.65 & 48 & 0.82 $\pm$ 0.08 \\
& & & 3.4 & 2.21 $\times$ 0.50 & 56.99 & 36 & 0.72 $\pm$ 0.06 \\
& & & 5 & 1.78 $\times$ 0.37 & 56.19 & 30 & 0.59 $\pm$ 0.05 \\
& & & 7 & 1.30 $\times$ 0.24 & 55.64 & 24 & 0.46 $\pm$ 0.04 \\
& & & 9 & 0.46 $\times$ 0.17 & 75.97 & 22 & 0.41 $\pm$ 0.04 \\
& & & 11 & 0.90 $\times$ 0.16 & 56.41 & 26 & 0.28 $\pm$ 0.04 \\ 
\hline
J1226$+$0815 & \ra{12}{26}{03.643} & \dec{+08}{15}{19.008} & 1.3 & 4.02 $\times$ 1.25 & 70.63 & 66 & 1.22 $\pm$ 0.11 \\
& & & 1.8 & 7.04 $\times$ 0.82 & 57.41 & 69 & 1.14 $\pm$ 0.11 \\
& & & 2.6 & 4.68 $\times$ 0.67 & 57.98 & 47 & 0.93 $\pm$ 0.08 \\
& & & 3.4 & 3.53 $\times$ 0.48 & 58.85 & 40 & 0.80 $\pm$ 0.07 \\
& & & 5 & 2.76 $\times$ 0.35 & 56.70 & 30 & 0.65 $\pm$ 0.05 \\
& & & 7 & 2.29 $\times$ 0.23 & 56.54 & 28 & 0.47 $\pm$ 0.05 \\
& & & 9 & 0.62 $\times$ 0.19 & 68.93 & 26 & 0.43 $\pm$ 0.04 \\
& & & 11 & 1.55 $\times$ 0.16 & 56.53 & 28 & 0.33 $\pm$ 0.04 \\ 
\hline
J1253$-$0312 & \ra{12}{53}{05.969} & \dec{-03}{12}{58.752} & 1.3 & 4.53 $\times$ 1.45 & 64.51 & 200 & 3.26 $\pm$ 0.26 \\
& & & 1.8 & 8.57 $\times$ 0.92 & 56.86 & 127 & 2.91 $\pm$ 0.19 \\
& & & 2.6 & 5.55 $\times$ 0.68 & 55.85 & 110 & 2.56 $\pm$ 0.17 \\
& & & 3.4 & 4.15 $\times$ 0.48 & 56.51 & 102 & 2.24 $\pm$ 0.15 \\
& & & 5 & 3.32 $\times$ 0.36 & 55.78 & 68 & 1.80 $\pm$ 0.11 \\
& & & 7 & 0.84 $\times$ 0.25 & 64.30 & 94 & 1.31 $\pm$ 0.11 \\
& & & 9 & 2.06 $\times$ 0.19 & 55.29 & 61 & 1.05 $\pm$ 0.08 \\
& & & 11 & 1.84 $\times$ 0.16 & 55.63 & 57 & 0.95 $\pm$ 0.07 \\ 
\enddata
\end{deluxetable*}

%\tablecomments{Optional\\}

\subsection{Additional Archival Data}\label{archival_data}
In conjunction with our VLA and EVN observations, we searched the positions of all sources in our sample in the LOFAR Two-metre Sky Survey (LoTSS) Data Release 2 (DR2), which consists of over 4 million radio sources at a central frequency of 144\,MHz \citep{LoTSS}. The LoTSS DR2 observations cover 27$\%$ of the northern sky with an angular resolution of $\sim$ 6\arcsec~and a median rms sensitivity of $\sim$ 80 $\upmu$Jy/beam. We found three out of 12 of our targets were located within the LoTSS footprint: J0903+4824, J0909+5655, and J0931+5633, with integrated flux densities of 36.4 $\pm$ 0.4 mJy, 2.2 $\pm$ 0.1 mJy, and 9.3 $\pm$ 0.3 mJy, respectively. The positions of the remaining nine sources were not covered by LoTSS.

%2014 April 23 to 2020 February 5

In addition, we included all detections from Epochs 1 through 3 of the the VLA Sky Survey (VLASS; \citealt{VLASS}) at 3\,GHz for each source. VLASS offers an angular resolution of 2.5\arcsec~and a sensitivity of $\approx$ 120\,$\upmu$Jy rms per epoch. We extracted the flux densities at the source positions from the Quick Look image products, and applied corrections of 10$\%$ and 3$\%$ to the flux uncertainties in the first and subsequent epochs, respectively, to account for a systematic inaccuracy for sources below $\approx$ 1~Jy \citep{lacy2019vlass}. The uncertainties on the flux densities reflect the quadrature sums of the flux error and the systematic error, as described above. 

Lastly, we included data from Faint Images of the Radio Sky at Twenty cm (FIRST; \citealt{FIRST}), the NRAO VLA Sky Survey (NVSS; \citealt{NVSS}), and archival VLA observations, including upper limits and detections, for J0106$+$0046, J1136$+$2643, J1200$-$0341, and J1220$+$3020, as originally compiled in \cite{Eftekhari20}. For the flux uncertainties, we adopt the rms values reported in the FIRST catalog, and include an additional 5$\%$ systematic error to account for the larger integrated flux density errors observed in bright sources \citep{FIRST}. It is important to note that our sample was also observed with the VLBA at 9\,GHz \citep{Sargent22}. Among the sources, J0019$+$1507, J0909+5655, and J1136+2643 were detected, while the rest of the sample yielded non-detections. The archival observations were used in Sections \ref{subsec:rnorm} -- \ref{subsec:LCs}.

\subsection{Single-pulse Searches}\label{sec:single-pulse}

In addition to standard continuum observations with the EVN, we utilized the most sensitive telescope in the EVN array, Effelsberg, to search for FRBs in the direction of our EVN-observed target J0019$+$1507 within the frequency range $1.6-1.7$ GHz. Employing the custom data search pipeline detailed in \cite{Kirsten2024}, we processed raw voltages into 8-bit Stokes I filterbank files with a time resolution of $0.128$ ms and a frequency resolution of $62.5$ kHz. We also applied a static mask to excise channels with known radio frequency interference (RFI). 

To identify burst candidates, we applied the \texttt{Heimdall}\footnote{ https://sourceforge.net/projects/heimdall-astro/} search algorithm with a signal-to-noise ratio (S/N) threshold of $7$, searching over a broad dispersion measure (DM) range of $10-1500$ \dmunit. To approximate a DM value for source J0019$+$1507, we assumed the combined Milky Way (MW) halo and host contribution to be 100 \dmunit and determined the Galactic interstellar medium (ISM) contribution to be $\approx$ 37 \dmunit with the NE2001 model \citep{Cordes02}. We inferred a DM value of $\approx$ 170 \dmunit, which is significantly lower than the maximum of the search range. Lastly, we classified and refined the burst candidates ($> 5000$) identified by \texttt{Heimdall} with a machine learning classifier, FETCH, using models A and H and a probability threshold of $0.5$ \citep{agarwal_2020_mnras}. Additionally, we manually inspected all \texttt{Heimdall} reported candidates and found them to be either RFI or false positives.

No pulses were detected in $\approx 203$ minutes of on-source integration time. Thus, we place an upper limit using the radiometer equation (Equation 15 in \citealt{Cordes_2003_ApJ}), with a S/N of $7$, two summed polarizations, $\Delta \nu = 128$~MHz, a nominal duration of $1$~ms for the burst width, and a typical Effelsberg system temperature of $20$~K and gain of $1.54$~Jy~$\textrm{K}^{-1}$, yielding a fluence of $<0.18$~Jy~ms.

\subsection{Optical Imaging}\label{sec:optical_image}

As discussed in \cite{Reines20}, the dwarf galaxy population is selected from the NASA-Sloan Atlas with available optical imaging from the Sloan Digital Sky Survey (SDSS; \citealt{Blanton17}). To ascertain the potential presence of background interlopers among our sources, we examined the full sample with the DESI Legacy Imaging Surveys Data Release 10 \citep{Dey2019}. We find no apparent optical sources at the locations of our sources except for source J1027$+$0112, which is coincident with an apparently red point-like source with a dereddened $r$- and $z$-band brightness of 23.96~AB~mag and 21.88~AB~mag, respectively. Based on the spatial coincidence and color, we conclude that source J1027$+$0112 is likely a background AGN unconnected to the dwarf galaxy. We thus remove the source from our sample as a potential FRB-PRS candidate, and instead use it as a source for comparison.

For sources J0019$+$1507 and J0909$+$5655 which are compact on milliarcsecond (mas) scales and highly offset from their host galaxies, we obtained deep ground-based observations with Binospec mounted on the 6.5~m MMT telescope (2023B-UAO-G201-23B; PI: A. Nugent) and DEIMOS on the 10~m Keck telescope (O438; PI: A. Gordon). For data reduction and stacking, we utilized the \texttt{POTPyRI} software\footnote{https://github.com/CIERA-Transients/POTPyRI}. We further confirm the absence of any obvious coincident optical source in $z$-band at the locations of sources J0019$+$1507 and J0909$+$5655.

\section{Analysis $\&$ Results} \label{sec:res}
In this section, we examine various properties of the 12 radio sources in our sample including their physical sizes, host-normalized offsets, SEDs, spectral luminosities, and light curves. We compare them to the known PRSs associated with FRBs: PRS\,20121102A \citep{Chatterjee17} and PRS\,20190520B \citep{Niu22} to assess the likelihood that they share a common origin. In our comprehensive analysis, we also include additional radio transients discovered within dwarf host galaxies including the decades-long transient J1419$+$3940 \citep{Law18}, the PRS associated with the SLSN PTF10hgi \citep{Eftekhari19}, and the PWN candidate VT\,1137$-$0337 (Epoch 2; \citealt{DDong23}). Noteworthy for their similar host environments, these transients also exhibit radio properties reminiscent of PRSs, hinting at a potential common origin with the PRSs associated with FRBs. We further include a sample of compact radio sources in nearby dwarf galaxies, as identified by \cite{Vohl23} with LOFAR, akin to our PRS population and those associated with an FRB, to serve as a point of comparison. Our main aim is to assess which sources are most plausibly associated with FRBs, and thus we refer to our sample as ``PRS candidates'' unless proven otherwise.

\subsection{Characteristics of PRSs} \label{subsec:PRS_characteristics}
We first summarize the key defining characteristics of PRSs to identify sources in our sample that share these properties. PRSs are typically compact continuum radio sources and, thus far, only two have been identified in association with FRBs \citep{Chatterjee17,Niu22}. A third PRS was recently claimed to be associated with the active FRB\,20201124A \citep{Bruni23}. However, the apparent offset of the PRS from the FRB location and a lack of definitive evidence as to its compact nature makes this association more tentative. 

The compact nature of PRSs\,20121102A and 20190520B was established through high-resolution VLBI radio observations, placing constraints on their physical sizes down to pc and sub-pc scales \citep{Marcote17, Bhandari+23b}. The two FRB-PRS pairs were notably discovered in dwarf host environments \citep{Chatterjee17, Niu22}. PRSs\,20121102A and 20190520B exhibit angular offsets from their host centers of 0.2\arcsec~and 1.3\arcsec, respectively, indicating that they do not reside directly in their galaxy's cores. Moreover, \cite{Law22} proposed a specific luminosity threshold of L$_{\nu}$ $>$ 10$^{29}$ ergs s$^{-1}$Hz$^{-1}$ for PRSs based on the two known FRB-PRSs, effectively excluding radio emission from ongoing star formation and individual SNRs that are significantly less luminous. However, given the uncertainty of the intrinsic luminosity function of PRSs, we do not enforce the proposed luminosity threshold in ruling out PRS candidates. Thus, we use the following observed properties of known FRB-PRSs as general guidelines to characterize the sources in our sample:

\begin{enumerate}
  \item compact on $\lesssim$~10~pc (mas) scales,
  \item projected offset from the center of the host galaxy of $<$ 2 r$_\mathrm{e}$
\end{enumerate}

\noindent where r$_\mathrm{e}$ is the half-light radius of the host galaxy. Here, we choose 2~r$_\mathrm{e}$ because the chance coincidence probability for a radio source is sufficiently low ($P_{\rm cc} \sim 10^{-5}$) within such a region \citep{Eftekhari2018}. Recognizing that some AGN contamination may persist in our sample (see Section \ref{sec:AGNorigin}), we can establish an upper limit on the fraction of the sources potentially associated with FRBs.

We subsequently divide our sample into three tiers based on their observed properties. From here on, we refer to sources that satisfy all of our above criteria as ``Tier A". Sources in ``Tier B" satisfy only the first criterion, whereas ``Tier C" sources do not satisfy the first criterion, i.e., are resolved out on mas scales. Based on our criteria, we identify a single source in Tier A (J1136$+$2643), two sources in Tier B (J0019$+$1507 and J0909$+$5655), and six sources in Tier C (J0106$+$0046, J0903$+$4824, J0931$+$5933, J1200$-$0341, J1226$+$0815, and J1253$-$0312).  We note that we do not include J1027$+$0112, J1136$+$1252 and J1220$+$3020 in our classification scheme given their classification as confirmed or likely AGN, but we nevertheless summarize their observed properties in the following sections as compact radio sources of known origin.

\begin{deluxetable*}{cccccccccc}[t!]
\linespread{1.2}
\tabletypesize{\scriptsize}
\tablecaption{Radio Source Comparisons \label{tab:bigdata}}
\tablecolumns{10}
\tablewidth{0pt}
\tablehead{
\colhead{Source} &
\colhead{z} &
\colhead{Size} & 
\colhead{Tier} &
\colhead{$r_{\mathrm{norm}}$} &
\colhead{$L_{\mathrm{3\,GHz}}$} &
\colhead{SED shape} &
\colhead{$\alpha$} & 
\colhead{$\nu_{\mathrm{b}}$} &
\colhead{References} \\
\colhead{} &
\colhead{} & 
\colhead{(pc)} & 
\colhead{} & 
\colhead{($r_{\mathrm{e}}$)} & 
\colhead{($\mathrm{ergs~s}^{-1}~\mathrm{Hz}^{-1}$)} &  
\colhead{} & 
\colhead{} & 
\colhead{(GHz)}}
\startdata
PRS\,20121102A & 0.1927 & [0.03, 0.7] & & 0.4 & $2.3 \times 10^{29}$ & BPL & [$-$0.07, $-$1.31] & $\sim$~9 & 1$-$5 \\
PRS\,20190520B & 0.241 & $<$9 & & 1.3 & $3.0 \times 10^{29}$ & SPL & $-$0.33 & $\ldots$ & 6$-$8 \\
VT 1137$-$0337 & 0.0276 & $<$6.1 & & 0.8 & $2.3 \times 10^{28}$ & SPL & $-$0.35 & $\ldots$ & 9 \\
PTF10hgi & 0.098 & $<$2060 & & $\ldots$ & $1.2 \times 10^{28}$ & SPL &  $-$0.14 & $\ldots$ & 10 \\
J1419$+$3940 & 0.0196 & 1.6 & & 0.7 & $9.5 \times 10^{27}$ & SPL & $\lesssim-0.65$ & $\ldots$ & 11, 12 \\
\hline
J0019+1507 & 0.0376 & $<$1.95 & B & 5.3 & $7.8 \times 10^{28}$ & SPL & $-$0.03 $\pm$ 0.02 & $\ldots$ & 13, This work \\
J0106$+$0046 & 0.0171 & [11, 87] & C & 0.6 & $6.3 \times 10^{27}$ & SPL & $-$0.92 $\pm$ 0.04 & $\ldots$ & 14, This work \\
J0903$+$4824 & 0.0272 & [17, 102] & C & 1.6 & $3.8 \times 10^{28}$ & SPL & $-$0.92 $\pm$ 0.01 & $\ldots$ & 14, This work \\
J0909$+$5655 & 0.0315 & $<$1.33 & B & 3.0 & $3.2 \times 10^{28}$ & BPL & [$-$0.18 $\pm$ 0.01, $-$0.73 $\pm$ 0.10] & 4.8 $\pm$ 0.6 & 13, 14, This work \\
J0931$+$5633 & 0.0494 & [30, 200] & C & 4.1 & $3.3 \times 10^{29}$ & BPL & [0.0 $\pm$ 0.01, $-$0.78 $\pm$ 0.03] & 1.4 $\pm$ 0.1 & 14, This work \\
J1136$+$2643 & 0.0331 & $<$1.43 & A & 1.3 & $5.3 \times 10^{28}$ & BPL & [0.77 $\pm$ 0.24, $-$1.65 $\pm$ 0.19] & 3.6 $\pm$ 0.2 & 13, 14, This work \\
J1200$-$0341 & 0.0257 & [17, 182] & C & 1.2 & $5.1 \times 10^{28}$ & SPL & $-$0.55 $\pm$ 0.02 & $\ldots$ & 14, This work \\
J1226$+$0815 & 0.0241 & [16, 111] & C & 0.0 & $1.2 \times 10^{28}$ & SPL & $-$0.59 $\pm$ 0.04 & $\ldots$ & 14, This work \\
J1253$-$0312 & 0.0221 & [14, 102] & C & 0.1 & $2.9 \times 10^{28}$ & BPL & [$-$0.35 $\pm$ 0.09, $-$0.83 $\pm$ 0.08] & 3.8 $\pm$ 0.6 & 14, This work \\
\hline
J1027$+$0112 & $\ldots$ & $\ldots$ & AGN(?) & $\ldots$ & $3.7 \times 10^{28}$ & SPL & $-$0.19 $\pm$ 0.03 & $\ldots$ & 14, This work \\
J1136$+$1252 & 0.76 & $<$9.86 & AGN & $\ldots$ & $7.2 \times 10^{31}$ & SPL & 0.12 $\pm$ 0.04 & $\ldots$ & 14, 15, This work \\
J1220$+$3020 & 0.0269 & [17, 101] & AGN & 0.1 & $1.4 \times 10^{28}$ & SPL & $-$0.51 $\pm$ 0.07 & $\ldots$ & 14, This work \\
\enddata 
\tablecomments{Radio-inferred properties of our 9 PRS candidates, 2 AGN and 1 likely AGN sources in our sample, PRSs associated with FRBs, and radio transients VT1137$-$0337, PTF10hgi, and J1419$+$3940. The SED shape is either a single power law (SPL) or a broken power law (BPL). Column 1: source name. Column 2: physical size from major axis diameter. Column 3: classification based on criteria discussed in Section \ref{subsec:PRS_characteristics}. Column 4: host-normalized offsets.  Column 5: spectral luminosity at 3\,GHz. Column 6: shape that best characterizes the SED. Column 7: best-fit spectral indices. Column 8: break frequency $\nu_{b}$ (if applicable). Column 9: references.
\textbf{References} 
(1) \citealt{Chatterjee17},
(2) \citealt{Marcote17}, 
(3) \citealt{Resmi21}, 
(4) \citealt{Tendulkar17}, 
(5) \citealt{Chen23},
(6) \citealt{Bhandari+23b},
(7) \citealt{Niu22},
(8) \citealt{Zhang23},
(9) \citealt{DDong23},
(10) \citealt{Eftekhari19},
(11) \citealt{Law18},
(12) \citealt{Marcote19},
(13) \citealt{Sargent22},
(14) \citealt{Reines20}.
(15) Sturm et al. 2024, \textit{in prep.}
}
\end{deluxetable*}

% J1027$+$0112 & $<$0.89

\subsection{Physical Sizes $\&$ Host-normalized Offsets}\label{subsec:rnorm}

We take advantage of our high-resolution observations to constrain the physical sizes of the \nsamp~PRS candidates in our sample under the assumption that they are at the redshifts of the dwarf galaxies. Beginning with the EVN-detected source, J0019$+$1507 (Figure~\ref{fig:EVN_sources}), we find that it is unresolved on mas scales, with corresponding constraints on its physical size of $<$ 3.5 pc. J0019$+$1507 is also detected with the VLBA at 9~GHz taken at a slightly higher resolution \citep{Sargent22}; we adopt the more constraining size limit of $<$ 1.95 pc for this source. Two additional sources (J0909$+$5655 and J1136$+$2643) are detected with the VLBA, with physical sizes of $<$ 1.33 pc and $<$ 1.43 pc, respectively \citep{Sargent22}. We therefore identify a total of three sources in our sample (J0019+1507, J0909+5655, and J1136$+$2643) that satisfy criteria \#1 in Section~\ref{subsec:PRS_characteristics} (i.e., compact on pc scales), with physical extents of $\approx$ 1.3 -- 2 pc. The physical sizes of these sources are comparable to those of both known PRSs ($<9$ pc), the pulsar wind nebula candidate VT 1137$-$0337 ($< 6.1$ pc), and the decades-long radio transient J1419$+$3940 (1.6 pc). 

Of the remaining six sources in our sample with VLA observations, none were observed with the EVN at 1.7\,GHz, and all six were non-detections with the VLBA \citep{Sargent22}. For these sources, we derive both upper and lower limits on the physical sizes of each source. Specifically, we calculate an upper limit from the VLA A-configuration observations using the major axes of the synthesized beams \citep{Reines20} and a lower limit from VLBA non-detections using the largest angular scale (LAS) of $\sim$ 30 mas at 9\,GHz \citep{Sargent22}. The larger upper limits of 87 -- 200~pc can be attributed to either the absence of high-resolution observations or an intrinsically larger physical size. Finally, we note that J0903$+$4824 (Tier C) is resolved in our VLA observations over 2 -- 12\,GHz (S-, C-, and X-bands), consistent with previous findings \citep{Reines20}. All derived physical sizes are calculated assuming the radio sources are at the redshift of their dwarf host galaxies, and are only valid under this condition. 

We note that in addition to the PRSs in our sample, we also detected the background source J1027$+$0112 and the background AGN J1136$+$1252 in our EVN observations. J1136$+$1252 has a physical size of $<$ 9.86 pc at a redshift of $z = 0.76$ (Sturm et al. 2024, \textit{in prep.}), indicating that the radio emission is core-dominated. Interestingly, J1136$+$1252 is not detected with VLBA at 9\,GHz. A potential explanation is refractive scintillation of the compact core due to small-scale irregularities in the ISM. Indeed, given the Galactic latitude of the source, variations in the flux density of up to 60$\%$ at higher frequencies are expected \citep{Walker98}. The scintillation level may be even higher depending on the observation time, as the timescale is modulated by Earth's orbit \citep{Cordes98}. We therefore attribute the VLBA non-detection to scintillation-induced variability. The non-detection of the AGN source J1220$+$3020 with VLBA indicates that the radio source is effectively ``resolved out'', exhibiting a physical extent beyond the LAS of the VLBA observations. We report the source sizes of our sample in Table \ref{tab:bigdata}.

Next, we consider the relative locations of these sources within their host galaxies. For sources detected with either the EVN or VLBA, we calculate the angular offsets from their host centers using these more precise positions. When unavailable, we adopt the values compiled in \cite{Reines20} using high-resolution VLA observations in A-configuration. We find that more than half of the sources exhibit spatial offsets of $\gtrsim$ 2\arcsec~from their host optical centers. Among the three VLBA-detected sources, J0909$+$5655 and J1136$+$2643 exhibit offsets $\lesssim$ 3\arcsec~from the photocenters of their host galaxies, while J0019+1507 is offset by $>$ 4\arcsec. The large offset is consistent with the off-nuclear locations of the FRB-PRSs \citep{Eftekhari20}. On the other hand, in the context of AGN \citep{Reines20}, the large spatial offsets may be attributable to the weaker gravitational potentials in less massive dwarf galaxies \citep{Shen19}, and/or high-velocity kicks caused by merging BHs. 

Despite sharing similar stellar masses, the dwarf hosts in our sample exhibit diverse galaxy morphologies and sizes (see Figure 7 in \citealt{Reines20}). To enable a fair comparison across our sample and provide a more standardized offset measurement, we therefore compute the host-normalized offset (r$_\mathrm{norm}$) for each source, in which the angular offset is normalized by r$_\mathrm{e}$. For the half-light radii of the dwarf galaxies, we utilize the semi-major axes reported NASA/IPAC Extragalactic Database (NED)\footnote{https://ned.ipac.caltech.edu/} using an exponential light profile.

\begin{figure}[t!]
    \centering
    \includegraphics[width=0.48\textwidth]{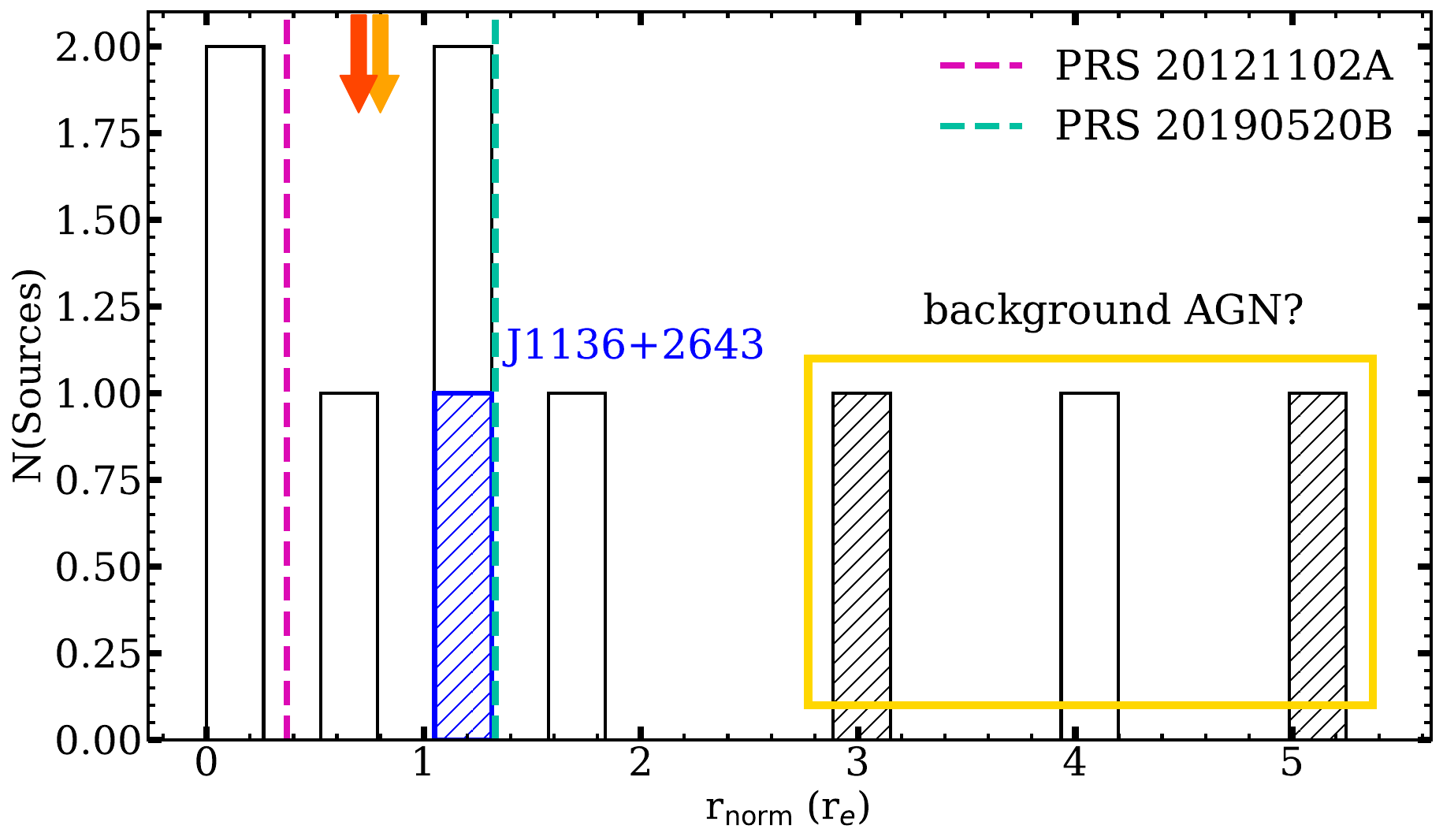}
    \caption{Host-normalized offset distribution of the \nsamp~radio sources in our sample. Hatched bins represent sources that are confirmed as compact on mas scales. Sources with r$_\mathrm{norm}$ values $\gtrsim$ 3~r$_\mathrm{e}$ are boxed in yellow as potential background AGN. 
    The magenta and green lines denote the the host-normalized offsets of PRSs\,20121102A and 20190520B, while the r$_\mathrm{norm}$ values of J1419$+$3940 and VT1137$-$0337 are indicated by the red and orange arrows, respectively. Based on its comparable physical size and spatial offset, we identify J1136$+$2643, represented by the blue hatched bin, as the most promising FRB-PRS candidate among the observed sources.} 
    \label{fig:offsets}
\end{figure}

For comparison, we employ the host-normalized offsets of the two known FRB-PRSs. For PRS\,20121102A, we adopt r$_\mathrm{norm}$ = 0.4~r$_\mathrm{e}$ \citep{Mannings21}.
For PRS\,20190520B, we derive r$_\mathrm{e}$ for the host galaxy by modeling the archival CFHT/MegaCam R-band imaging with \textsc{Galfit} (v3.0.5; \citealt{galfit_2002,galfit_2010}). We first model the empirical point spread function (ePSF) of the image using the \texttt{EPSFBuilder} module of \textsc{photutils} (v0.6; \citealt{photutils}). Next, we initialize \textsc{Galfit} with the ePSF and a single S\'ersic surface brightness profile \citep{Sersic_1968} at the position of the host with a fixed S\'ersic index of n = 1 and position angle of $-30^{\circ}$ \footnote{We note that due to the faintness of the galaxy, the fit does not converge on reasonable values if we let the S\'ersic index be free.}, leaving all other default parameters as free. The resulting best-fit model yields r$_\mathrm{e}$ = 0.98\arcsec, corresponding to a host-normalized offset r$_\mathrm{norm}$ = 1.3~r$_\mathrm{e}$ (for an angular offset of 1.3\arcsec~; \citealt{Niu22}). We also include in Figure \ref{fig:offsets} the host-normalized offsets for J1419$+$3940 \citep{Law18} and VT1137$-$0337 \citep{DDong23}, where we adopt angular offsets of 0.5~pc and 0.4~pc, respectively, and normalize them by the sizes of their hosts as reported in NED. We exclude PTF10hgi from this analysis due to the lack of a reported host size for this source. 

In Figure \ref{fig:offsets}, we plot the host-normalized offset distribution for all \nsamp~sources. We find a wide range of r$_\mathrm{norm} \approx 0~$\textendash$~5.3~\rm r_{e}$ with upper and lower bounds set by J0019$+$1507 and J1226$+$0815, respectively, and a median of 1.3 r$_\mathrm{e}$. Among the entire population, J1136$+$2643 (denoted by the blue hatched bin) is the only source that shares both a similar host-normalized offset and compact physical size with the FRB-PRSs. Three sources (J0019$+$1507, J0909$+$5655, and J0931$+$5633) exhibit r$_\mathrm{norm}$ values $\gtrsim 3~\rm r_e$, larger than those of PRSs\,20121102A and 20190520B by at least a factor of 2. Two of these highly offset sources are confirmed as compact with VLBI as represented by the black hatched bins in Figure~\ref{fig:offsets}. The large spatial offsets observed for these sources indicate a likely background origin \citep{Sargent22}. In contrast, sources J0106+0046, J1200$-$0341, J1226$+$0815 and J1253$-$0312 in Tier C are located closer to the host nucleus with r$_\mathrm{norm}$ values comparable to both FRB-PRSs as well as transients J1419$+$3940 (0.7~r$_\mathrm{e}$) and VT1137$-$0337 (0.8~r$_\mathrm{e}$). As alluded to in Section \ref{subsec:PRS_characteristics}, the potential heterogeneity in our sample means that the sources closest to the nucleus may represent a subset of underlying AGN. Finally, while the sample of compact radio sources presented by \cite{Vohl23} predominantly (16 out of 27 sources) exhibit an angular offset of $\lesssim$ 2\arcsec~due to the nature of selection, their host-normalized offsets are not available for comparison.

Based on our selection criteria as defined in Section \ref{subsec:PRS_characteristics}, we therefore find J1136$+$2643 to be the most promising analog to PRSs associated with FRBs. Conversely, although J0019+1507 and J0909+5655 are compact on similar spatial scales, they exhibit large host-normalized offsets and their origin as background sources cannot be ruled out. Thus, these sources are relegated to Tier B. Individual values for r$_\mathrm{norm}$ are provided in Table \ref{tab:bigdata}.

\subsection{Spectral Energy Distributions} \label{subsec:SEDs}

\begin{figure*}[t!]
    \centering
    \includegraphics[width=\textwidth]{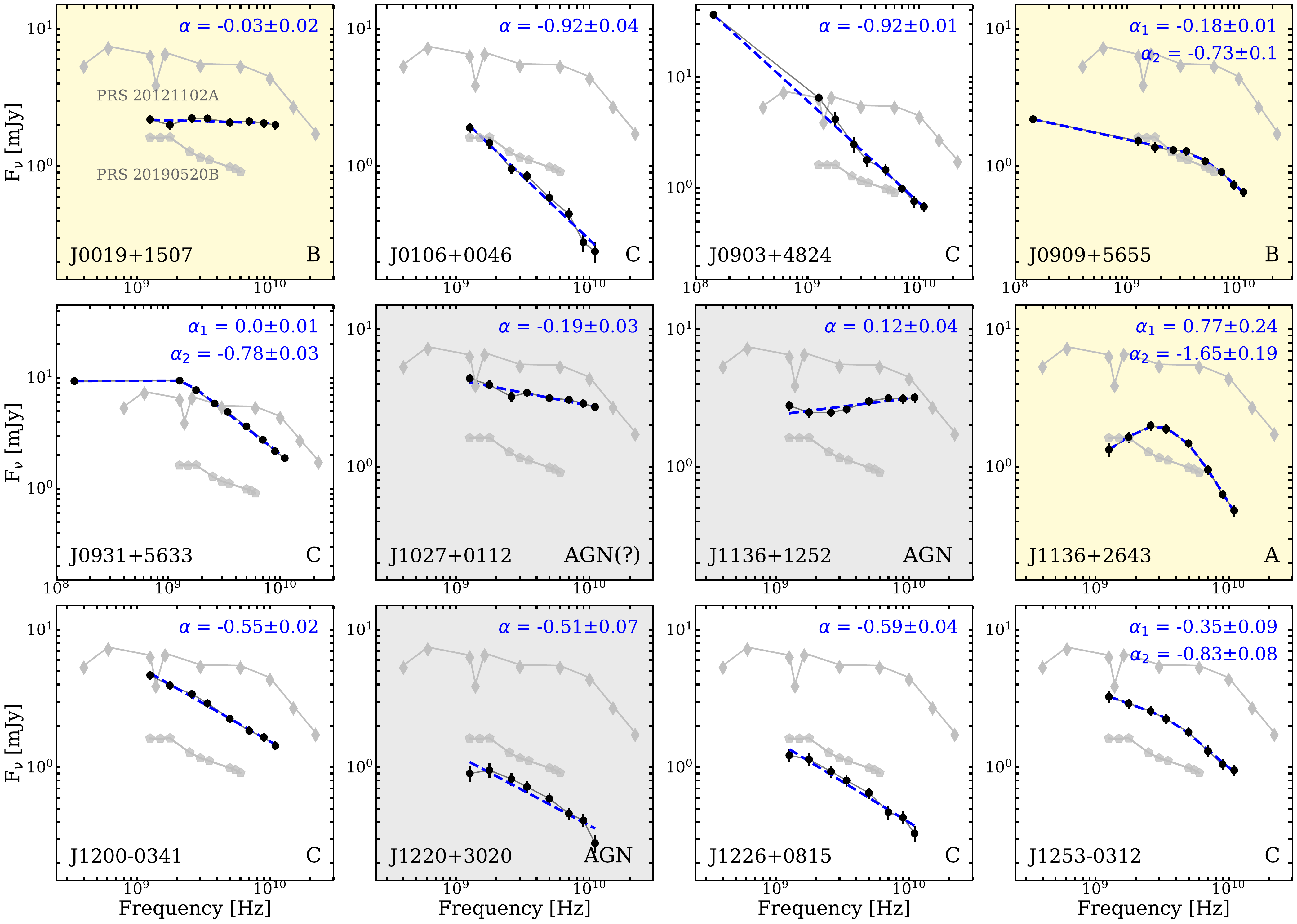}
    \caption{SEDs for the \nsamp~PRS candidates, 2 confirmed AGN and 1 likely AGN in our sample using flux densities from our VLA observations (1$-$12 GHz). We also include flux densities at 144\,MHz for the three sources detected in LoTSS. Blue dashed lines correspond to power law fits with the derived spectral indices and $1\sigma$ uncertainties listed in each panel. Yellow shaded panels denote sources that are compact on mas scales. The confirmed and possible AGN are highlighted in grey. The letter in the bottom right corner in each panel indicates the tier to which each source belongs based on our selection criteria. For comparison, we also show the SEDs of the PRSs associated with FRB\,20121102A (diamonds) and FRB\,20190520B (pentagons) with an arbitrary flux scaling. The SED of the Tier A source J1136$+$2643 exhibits a turnover at 3.6\,GHz and shape that is indicative of SSA, in agreement with its compact size. Sources in Tier B and C are characterized by both simple and broken power laws with spectral indices ranging from 0.12 to $-$0.92, demonstrating a diversity in the spectral shape.}
    \label{fig:SED}
\end{figure*}

In Figure~\ref{fig:SED}, we plot the radio SEDs spanning 1--12\,GHz for all \nsamp~sources, including the two confirmed AGN and the one likely AGN, using the VLA observations detailed in Section \ref{sec:obs}. We complement the SEDs for sources J0903$+$4824, J0909$+$5655, and J0931$+$5633 with LoTSS data at 144\,MHz with a temporal separation with the VLA data of up to $\sim$ two years, which we consider negligible given the apparent lack of variability in the light curves of radio transients at $<1$ GHz on these timescales \citep{Bell14, Rowlinson16}. For comparison, we plot the SEDs of PRSs\,20121102A and 20190520B with an arbitrary flux scaling in each panel. In particular, we construct the SED of PRS\,20121102A with data from the Giant Metre-wave Radio Telescope (GMRT) between 400\,MHz and 1.4\,GHz \citep{Resmi21} and the VLA between 1.6 and 22\,GHz \citep{Chatterjee17}. Similarly, we plot the radio SED of PRS\,20190520B using VLA data between 1.3 and 6\,GHz \citep{Niu22}.

To characterize the shape of the SEDs, we employ the \texttt{curve$\_$fit} function within the \texttt{SciPy.optimize} package \citep{scipy} and fit the data with a single power law, $F_\nu \propto \nu^{\alpha}$. For sources that exhibit a break in the SED, we adopt a smoothed broken power law of the form:

\begin{equation}
    F_\nu = C \left[ \left[\frac{\nu}{\nu_{b}} \right]^{-s_1\alpha_1} + \left [\frac{\nu}{\nu_{b}} \right]^{-s_1\alpha_2}  \right ]^{-1/s_1}
\end{equation}

\noindent where C is the normalization constant, $\nu_b$ is the break frequency, $\alpha_1$ and $\alpha_2$ are the spectral indices before and after the break, respectively, and $s_1$ is the smoothing parameter. Individual fits and best-fit parameters are shown in Figure \ref{fig:SED} and listed in Table \ref{tab:bigdata}.

% Tier A source discussion
We find that our Tier A source J1136$+$2643 is best characterized by a broken power law that becomes optically thin above a break frequency of $\nu_b$ = 3.6 $\pm$ 0.2 \,GHz. Such a spectral shape is a hallmark of synchrotron self-absorption (SSA; \citealt{Condon16}) which is indicative of a compact emission region. We use the observed turnover frequency to infer a physical size for the source in Section \ref{subsec:equipartition}. Interestingly, the SED of PRS\,20121102A also features a break in which it steepens from $\sim\nu^{-0.1}$ to $\sim\nu^{-1.3}$ at $\sim$ 9\,GHz \citep{Resmi21}, possibly indicative of synchrotron cooling. In contrast, the absence of a clear SSA feature in the SEDs of PRSs\,20121102A and 20190520B implies that such a turnover is located at frequencies below $\lesssim 1$ GHz. 

% Tier B discussion
Among our two Tier B sources, the SED for J0019$+$1507 is fairly flat across 1 -- 12 GHz with spectral index of $\alpha = -0.03$ while J0909$+$5655 displays a flat spectrum ($\alpha = -0.18$) from $1-4.8$ GHz followed by a steepening with $\alpha = -0.73$ at $\nu_{b}$ $\gtrsim 4.8$\,GHz. Such SEDs are broadly consistent with the flat spectral shapes observed in PWNe, generally characterized by spectral indices flatter than $-$0.5 \citep{Gaensler06}. Indeed, such a model has been proposed for PRS\,20121102A \citep{Dai2017, Kashiyama17, Yang19}. Furthermore, the flat spectral indices cannot be explained by diffusive shock interactions (DSA; \citealt{Bell78,Blandford87}), which typically yield steep spectral indices with $\alpha \lesssim -$0.5, in line with most synchrotron transients (e.g., \citealt{Chevalier11, Margutti19, Fang20}). Among our sample, we also find that the likely AGN J1027$+$0112 and background AGN J1136$+$1252 are the only other sources exhibiting a flat SED. 

% Tier C discussion
Finally, we find that all of the SEDs in our remaining six Tier C sources are characterized by fairly steep power laws with $\alpha \geq -0.92$. As discussed in Section \ref{subsec:rnorm}, source J0903$+$4824 is resolved in our VLA observations at higher frequencies and in VLA X-band observations from \cite{Reines20}. The slight feature between $\sim$ 1 and 3\,GHz in the SED  is therefore likely a result of two distinct components. We note that although sources J1200$-$0341, J1226$+$0815, and J1253$-$0312 exhibit spectral indices similar to PRS\,20190520B across the frequency range $\sim 1 - 3$\,GHz, they do not show evidence for flattening at $\lesssim$ 2 GHz. Considering the diversity in the spectral shapes observed across all tiers and their notable similarities to the known FRB-PRSs, we do not exclude these sources as plausible PRS candidates solely based on their SEDs.

\begin{figure}[hbt!]
    \centering
    \includegraphics[width=0.48\textwidth]{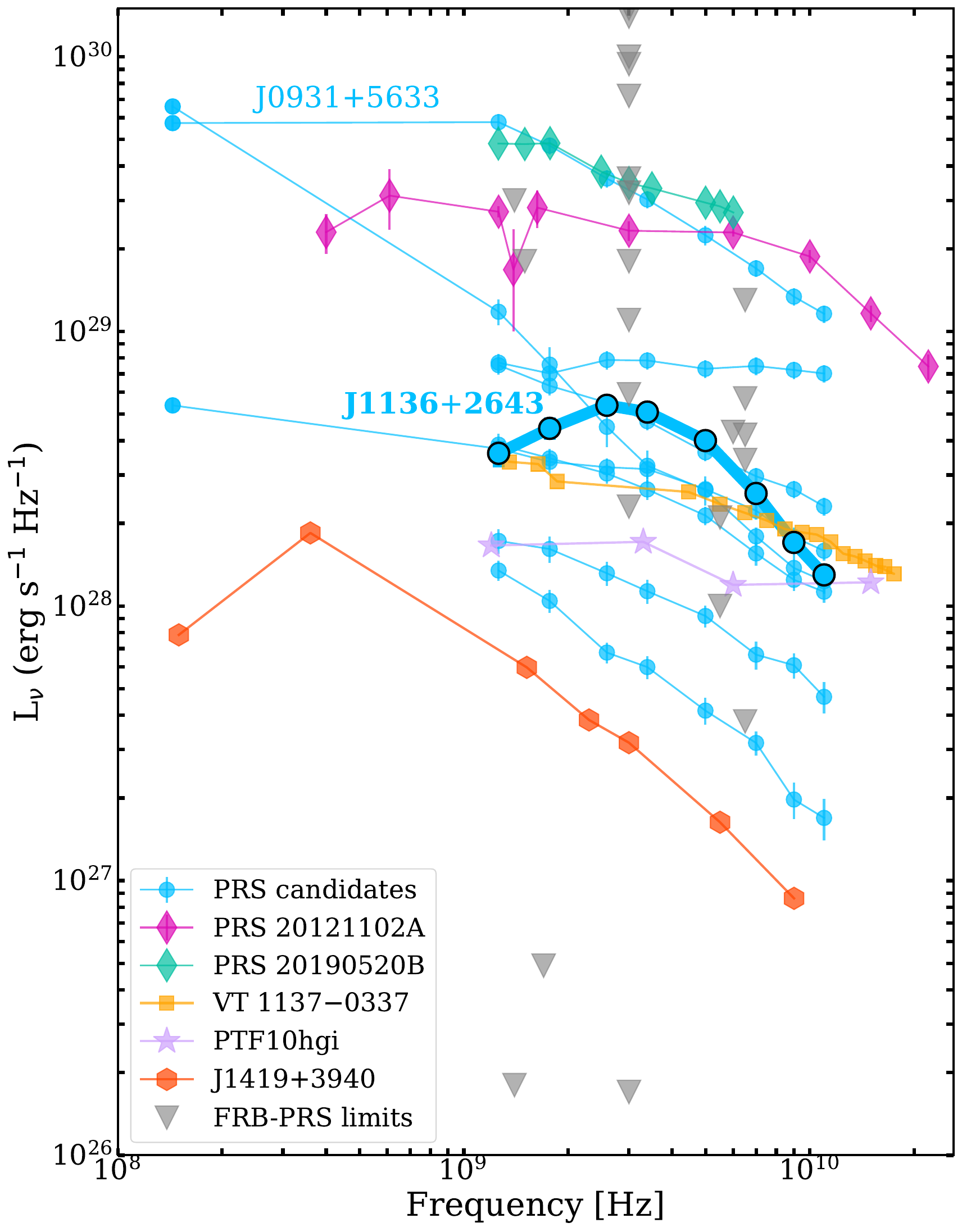}
    \caption{Radio spectral luminosities of the sample of PRS candidates (blue), where we highlight our Tier A source, J1136$+$2643. Also shown for comparison are PRS\,20121102A (pink; \citealt{Chatterjee17, Resmi21}), PRS\,20190520B (green; \citealt{Niu22, Zhang23}), VT 1137$-$0337 (orange; \citealt{DDong23}), PTF10hgi (purple; \citealt{Mondal20}), and J1419$+$3940 (dark orange; \citealt{Mooley22}). Grey triangles represent all existing PRS luminosity upper limits \citep{Law22, Law23}. The radio sources in our sample are overall less luminous than the FRB-PRSs, with the exception of J0931$+$5633, but comparable to other radio transients.}
    \label{fig:LSED}
\end{figure}

\subsection{Radio Luminosities}\label{sec:L_spec}

As shown in Table \ref{tab:bigdata}, the sources in our sample, along with other known PRSs, occupy a wide range of radio luminosities at 3\,GHz. Motivated by the diversity in luminosity and SED shapes (Section~\ref{subsec:SEDs}), we plot the spectral luminosities of the \nsamp~PRS candidates at the redshifts of the dwarf galaxies in comparison to those of FRB-PRSs, including limits on PRSs associated with FRBs \citep{Law22, Law23}, as well as known transients VT 1137$-$0337, PTF10hgi, and J1419$+$3940 in Figure~\ref{fig:LSED}. We compile the spectral luminosities of VT 1137$-$0337 from VLA observations taken in 2019 (Epoch 2 in \citealt{DDong23}) over $\sim$ 1 -- 18\,GHz. Similarly, the broad-band SED of the radio source associated with PTF10hgi is plotted utilizing VLA data from 2020 across the range $\sim$ 1 -- 15 GHz \citep{Mondal20}. Lastly, the SED for J1419$+$3940 is based on radio observations obtained in 2019 with LOFAR, VLA, and VLBA spanning 150\,MHz -- 9\,GHz \citep{Mooley22}. These specific epochs provide the broadest, contemporaneous spectrum for each radio transient.

The \nsamp~PRS candidates in our sample span about three orders of magnitude in luminosity, ranging from $\approx$ 10$^{27}$ to 10$^{30}$ erg s$^{-1} $ Hz$^{-1}$. While the Tier A source, J1136$+$2643, highlighted in Figure \ref{fig:LSED}, is less luminous than both PRSs known to be associated with FRBs, it does not significantly differ from those of the other sources in our sample and is comparable to the radio luminosity of VT 1137$-$0337. When compared to the known FRB-PRSs, all but J0931$+$5633 (Tier C) are less luminous by at least a factor of 10 (above 1\,GHz). Only two radio sources at the lower end of the luminosity distribution are dimmer than PTF10hgi, and none are less luminous than the decades-long transient J1419$+$3940. Given that only one source in our sample matches the luminosity of known FRB-PRSs, this suggests that the latter may be unusually luminous among the FRB-PRS population.

%discuss limits separately here.
In terms of the PRS limits associated with FRBs, a majority of them are shallower than our radio sources with only three being deeper by $\sim$ an order of magnitude. We also note that most of these limits are lower than those of PRSs\,20121102A and 20190520B. This supports the notion that if these PRSs exist, they exhibit lower luminosities, possibly similar to what is observed for our sources. Lower luminosities could easily be explained by factors such as older age or intrinsic lower energy output \citep{Margalit19}. 

As FRB-PRS candidates, we conclude that our sample of sources span a wide spectral luminosity range in which only the brightest source (J0909$+$5633, Tier C) is comparable to the luminosities of the FRB-PRSs. The remaining sources, including J1136$+$2643 in Tier A, are notably less luminous but in agreement with the spectral luminosities of VT 1137$-$0337 and PTF10ghi and more luminous than J1419$+$3940. Compared to the PRS limits, our sample is deeper than the majority of limits, with the exception of three deep limits at $\lesssim 10^{27} \ \rm erg \ s^{-1} \ Hz^{-1}$. Overall, we cannot rule out any of the sources in our sample as FRB-PRS candidates in the phase space of radio luminosities alone. More importantly, this highlights the presence of sub-luminous radio sources relative to the known FRB-PRSs in dwarf galaxies, most of which would have been ruled out by the PRS luminosity threshold proposed by \cite{Law22}.

\begin{figure*}
    \centering
    \includegraphics[width=0.7\textwidth]{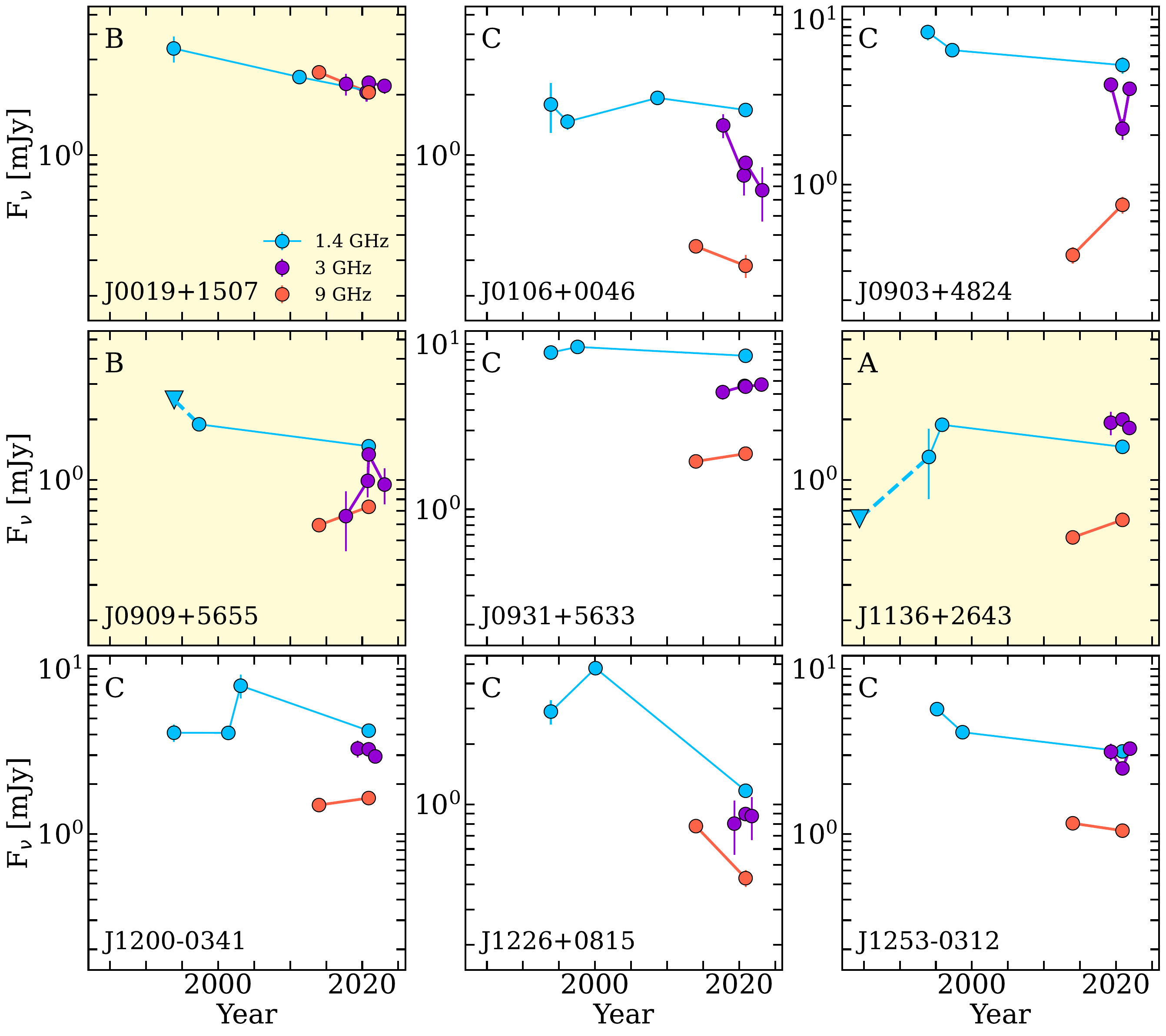}
    \caption{Radio light curves of the PRS candidates shown at 1.4 (blue), 3 (purple), and 9\, GHz (orange). The light curves at 1.4 GHz are drawn from FIRST, NVSS, and archival VLA data as detailed in \cite{Eftekhari20}, and our own VLA data at 1.4\,GHz. The panels highlighted in yellow are Tier A and B sources compact on mas scales. Data at 3 and 9\,GHz include observations from VLASS and \cite{Reines20}, in addition to our VLA observations. The 1.4\,GHz light curves cover the longest baseline, spanning approximately three decades, while the 3 and 9\,GHz light curves illustrate the temporal evolution over a shorter span of approximately 5$-$10 years. Detections are shown as circles, while upper limits are represented as triangles. Overall, the light curves at 1.4\,GHz are generally flat or gradually declining, including Tier A source J1136$+$2643, over a span of 3 decades. On shorter timescales, most light curves at 3 and 9\,GHz remain flat with minor variability. Due to the sparse data coverage and inhomogeneous comparison timescales, we do not eliminate any source as a PRS candidate. }
    \label{fig:LCs}
\end{figure*}

\subsection{Light Curves} \label{subsec:LCs}
The precise localizations of FRBs\,20121102A and 20190520B have facilitated continued monitoring campaigns to characterize the temporal evolution of their associated PRSs over various timescales, spanning from days/weeks to years. To explore whether the \nsamp~PRS candidates in our sample exhibit temporal variability similar to the PRSs associated with FRBs, we compile the light curves at 1.4, 3, and 9\,GHz using the radio observations presented in Sections \ref{sec:VLA_obs} and \ref{archival_data}. We also include archival upper limits and detections for sources J0106$+$0046, J1136$+$2643, and J1200$-$0341 at 1.4 GHz as compiled in \cite{Eftekhari20} along with VLA data at 9\,GHz from \cite{Reines20}. The light curves for all \nsamp~PRS candidates are shown in blue (1.4\,GHz), purple (3\,GHz), and orange (9\,GHz) in Figure \ref{fig:LCs}.

Based on prevailing FRB progenitor models (see Section \ref{sec:models}), PRSs coincident with FRBs are expected to evolve on timescales of $\sim$ decades to centuries following the SN explosion that gives rise to a compact central engine \citep{Margalit18, Sridhar+22}. Thus motivated, we first inspect the light curves at 1.4\,GHz which span the longest temporal baseline of $\sim$ 3 decades. To first order, the compact source J1136$+$2643 in Tier A exhibits a slight increase by a factor of 1.5 at early times followed by a gradual decrease in flux. In Tier B, sources J0019$+$1507 and J0906$+$5655 demonstrate a monotonic decline with a flux reduction of $\approx$ 40 and 20$\%$ spanning roughly 2 decades, respectively. The upper limit in the light curve of J0906$+$5655 does not preclude a rise at early times. In Tier C, we find that the light curves of sources J0909$+$4824 and J1253$-$0312 depict a markedly similar secular decline in flux. This is in contrast to sources J0106$+$0046, J1200$-$0341, and J1226$+$0815 which all display some level of variability in the light curves. Specifically, source J1226$+$0815 exhibits the most drastic decrease in flux of $\sim$ 75$\%$ over a $\sim 2$-decade timescale.

Compared to J1419$+$3940, the decades-long radio transient that decays to half of its peak flux value in 3 years \citep{Law18}, our source population is fading more gradually and therefore less energetic and relativistic.
%is less luminous by a factor of 10 at early times. 
We stress that such comparisons on $\sim$ decade timescales cannot be extended to the FRB-PRSs; due to their recent discovery dates, constraints on their long-term variability are not yet tenable. We note that neither of the FRB-PRSs is detected in FIRST or NVSS, with 3$\sigma$ limits of 3~mJy and 7.5~mJy, respectively.

To assess the level of variability for the PRS candidates on shorter $\sim 5$ year timescales, we turn to the light curves at 3 and 9\,GHz. In Tier A, the light curve of J1136$+$2643 does not show variability and remains relatively constant at 3\,GHz. However, a minor increase of 20$\%$ is observed at 9\,GHz. Furthermore, the light curve of Tier B source J0019$+$1507 remains relatively constant with a moderate variability of $\lesssim$ 20$\%$. In contrast, source J0909$+$5655 demonstrates a notable rise by a factor of 2 over a span of approximately 3 years, followed by a drop in the flux by a factor of 1.5. This rise and decline are reminiscent of the flux measurements of PRS\,20121102A observed on a time baseline of $\sim 7$ years with MeerKAT which revealed an increase of $\sim$ 30$\%$ and then a similar decline over a 3-year period, based on one epoch so far \citep{Rhodes23}. In this case, the observed temporal variability in PRS\,20121102A is likely intrinsic to the system (e.g., due to the interaction between a PWN terminal shock and the surrounding supernova ejecta; \citealt{Rhodes23}). However, it is worth mentioning that the observed variabilities in the known FRB-PRSs are not yet well established and that observed trends are based on limited time sampling. For compact sources, diffractive scintillation caused by ISM inhomogeneities is expected on timescales of hours to days, which are not probed by the light curves. However, refractive scintillation can induce slow, yearly variabilities in the light curve that is intrinsic to the source, as in the case of PRS\,20190520B \citep{Zhang23}. Due to limited time sampling, factors such as scintillation also cannot be ruled out as the source of observed variability with present data alone.

In Tier C, the short-term light curves of sources J0106$+$0046 and J1200$-$0341 exhibit a secular decline at 3\,GHz, analogous to the decrease in luminosity (20\%) observed for PRS\,20190520B from a multi-frequency (1 -- 12\,GHz) monitoring campaign between 2020 and 2021, alongside a 3.2$\sigma$ flux decrease at 3\,GHz \citep{Zhang23}. In contrast, the light curve of source J0931$+$5633 at 3\,GHz is characterized by an apparent, continual rise. While this trend could still be consistent with the flux increase observed in the light curve of PRS\,20121102A, it presents a challenge for scenarios such as the PWN model in which the flux is expected to decrease over time \citep{Yang16, Dai2017}

Finally, at 9\,GHz, the light curve of the Tier A source J1136$+$2643 is marginally rising over a $\sim$ 7 year period, in contrast to the declining behavior at lower frequencies. None of the sources in Tier B show evidence of substantial variability and out of the sources in Tier C, notable cases are J0903$+$4824, which exhibits a significant increase of $\sim$ 50$\%$, while J1226$+$0815 shows a decline of $\sim$ 50$\%$ over the span of 6.5 years. 

In conclusion, on a temporal baseline of $\lesssim 3$~decades, the light curves at 1.4\,GHz are generally flat or gradually decaying across all sources. At 3 and 9\,GHz, most light curves remain relatively constant with minor variability, comparable to the observed behaviors in the light curves of the known PRSs. Given the sparse and uneven data coverage across all frequencies, the diversity of PRS light curve behavior, and the inhomogeneous comparison timescales, the light curves at present do not offer obvious distinguishing power in determining the origin(s) of the radio sources in our sample. If most of the radio sources in our sample are in fact FRB-PRSs, this demonstrates diverse behavior across timescales and frequencies that cannot be easily explained with a single model.

\subsection{Coincident FRBs}\label{sec:coincidentFRBs}

To identify potential FRBs coincident with the locations of the PRS candidates in our sample, we utilized the Transient Name Server (TNS)\footnote{https://www.wis-tns.org/} to search for FRBs detected in the CHIME/FRB catalog \citep{CHIME_catalog} as well as FRBs detected by the by the Commensal Real-Time ASKAP Fast-Transients (CRAFT) survey \citep{CRAFT}. Our search did not yield any results from other experiments. For the CHIME/FRB catalog, we adopt a conservative search radius of $3^\circ$ centered around each radio source to accommodate the large sky localizations of CHIME FRBs and the non-Gaussian nature of the localization regions. For CRAFT, we impose a smaller search radius of 10\arcsec~to match the largest CRAFT FRB localization uncertainty (6\arcsec; \citealt{Prochaska2019}). Our initial search yielded a total of four unique, non-repeating CHIME FRBs coincident with three PRS candidates in our sample. 
Using the 10 GHz source counts from \cite{Mancuso2017}, we note, however, that the chance-coincidence probability for a $\sim 1$ mJy radio source in a $3^\circ$ radius region is $\sim 1.0$ \citep{Eftekhari2018}. Conversely, we find no CRAFT FRBs co-located with the sources in our sample, but note that the chance-coincidence probability in such a case would be $\sim$ 0.001.

To account for the non-Gaussian nature of the CHIME/FRB localization regions, we next manually check whether the matched sources fall within the localization regions for each FRB using the HDF5 localization files \citep{CHIME_catalog} and the CHIME/FRB Open Data Python package\footnote{https://chime-frb-open-data.github.io/}. We find that only one source, J0106$+$0046 (Tier C), falls within the 95$\%$ confidence interval of the localization region for FRB\,20190531E, while the rest are not spatially coincident with the FRB localization regions. 

To assess whether source J0106$+$0046 and FRB\,20190531E are plausibly related, we estimate the redshift probability distribution for FRB\,20190531E based on its DM of 328.2 pc cm$^{-3}$ \citep{CHIME_catalog}. Following the Macquart relation \citep{Macquart20}, we calculate the Galactic ISM contribution to be $\approx 33$ \dmunit from NE2001 \citep{Cordes02} and assume the DM contribution from the MW halo and the host to be 100 \dmunit. We find a redshift range for a confidence interval of 2.5$\%$ and 97.5$\%$ to be z = [0.12, 0.38] with a mean value of z = 0.28. This is well above the spectroscopic redshift of the dwarf host galaxy (z $\approx$ 0.02) associated with J0106$+$0046. Thus, if a plausible connection between source J0106$+$0046 and FRB\,20190531E exists, this requires an excess DM contribution (e.g., from the host) of $\approx$ 190 \dmunit. While this is not out of the question given other excess DM FRBs (e.g., an extreme case such as FRB\,20190520B with a foreground galaxy cluster contribution up to $\sim$ 640 pc cm$^{-3}$; \citealt{Lee23}), the majority of FRBs with known redshift do not exhibit such strong excess DM. Thus, we cannot conclusively associate any of the PRS candidates with coincident FRBs.

\section{FRB Progenitor Models} \label{sec:models}
At present, the emission mechanisms responsible for producing both FRBs and their associated PRSs are uncertain. A multitude of theoretical models have been proposed, typically invoking a compact neutron star as the central engine that powers a synchrotron nebula leading to the observed PRS emission \citep{Yang16, Kashiyama17, Margalit18, Li20}. In this section, we explore a few FRB progenitor models known to produce PRSs and test the viability of these models for our Tier A source J1136$+$2643, the strongest PRS candidate in our sample.

\subsection{Magnetar and Pulsar Wind Nebulae} \label{subsec:windneb}
The magnetar wind nebula model was first invoked to explain the observed properties (e.g., size and flux) of the PRS emission associated with FRB\,20121102A in which the non-thermal radio emission is powered by a magnetized nebula inflated by the flaring magnetar and confined behind the ejecta shell \citep{Beloborodov17, Margalit18, Zhao21}. The formation channel of the magnetar, whether prompt or delayed, dictates the nature of the ejecta, which can arise from the core collapse of a massive star (and which could potentially be associated with a SLSN/LGRB), binary neutron star (BNS) merger, or the accretion-induced collapse (AIC) of a white dwarf \citep{Margalit19}.

We adopt the analytic approach of \cite{Margalit19} for the expanding nebula in which the nebula luminosity depends on the energy injection rate of the magnetar $\dot{E}$ and the density of the surrounding medium. We use $\zeta \equiv $ M$_{\mathrm{ej}}$/v$^3_{\mathrm{ej}}$ as a proxy for the ejecta density ($\rho_{\mathrm{ej}} \sim \zeta t^{-3} $) where M$_{\mathrm{ej}}$ and v$_{\mathrm{ej}}$ are the total mass and mean velocity of the expanding ejecta, respectively. The primary difference between the SLSN/LGRB, BNS and AIC scenarios is the lower M$_{\mathrm {ej}}$ and higher v$_{\mathrm{ej}}$ in the latter two cases which sets a shorter free-free transparency time. To satisfy the observed peak luminosity $\simeq$ 5.4 $\times$ 10$^{28}$ erg~s$^{-1}$ Hz$^{-1}$ of source J1136$+$2643 at the break frequency $\nu_{b}$ = 3.6\,GHz, the energy injection rate must satisfy: 

\begin{equation}
    \dot E \simeq 2.15 \times 10^{41}~\mathrm{erg \, s^{-1}} \times \zeta_{-5}^{-41/144}
 \end{equation}

\noindent based on Equation 23 in \cite{Margalit19} where $\zeta \equiv 10^{-5} \zeta_{-5} \times (M_{\rm ej}/M_\odot) (v_{\rm ej}/10^4\,{\rm km \, s}^{-1})^{-3}$, and we have assumed that the mean energy per ion ejected in the magnetar wind and the wind magnetization are $\chi$ = 0.2\,GeV and $\sigma$ = 0.1, respectively. For these fiducial parameters, this implies an internal magnetic field for the putative magnetar of $\approx$ 2 $\times$ 10$^{16}$~G for a normal-mass NS within an expanding nebula that is 100 years old. (see Equation 2 of \citealt{Margalit19}). This is consistent with the inferred values for both PRSs\,20121102A and 20190520B ($\approx$ 10$^{16}$~G). Such a strong field points to a highly magnetized magnetar as a plausible central engine that could support the persistent radio emission on the observed timescale of decades for source J1136$+$2643.  

The observational constraints of J1136$+$2643 also allow us to rule out a significant portion of the $\dot{E}$-$\zeta$ phase space that corresponds to different progenitor formation channels as shown in Figure 4 of \cite{Margalit19}. In particular, we find that magnetars born from core-collapse SN and BNS mergers remains plausible. In the first case, a higher ejecta density necessitates a $\dot{E}$ value between 10$^{39-40}$ erg s$^{-1}$ whereas the lower $\zeta$ in the BNS merger scenario permits a rate of energy injection in the range of 10$^{41-42}$ erg s$^{-1}$. Lastly, we rule out the AIC scenario as it occupies the region where $\dot{E}$ and $\zeta$ values are both low and is thus inconsistent with the observed spectral luminosity of J1136$+$2643. 

If instead, the energy deposited into the wind nebula is powered by the spin-down of a pulsar, as in the case of a pulsar wind nebula \citep{Dai2017, Kashiyama17, Yang19}, we infer an external dipole field of $\approx$ 5 $\times$ 10$^{12}$ G assuming a central ionizing source equal to the spin-down luminosity of the pulsar. This inferred value is expected for an ordinary pulsar with a moderate spin period (P $\sim$ 0.1 -- 1 s, \citealt{Gaensler06}). Additionally, this poloidal field strength is improbable for the LGRB scenario as it falls short of that demanded by the spin-down luminosity from the relativistic jet of a LGRB which is usually much stronger (10$^{15}$ G) for a rapid spin period of P $\sim$ 1 -- 2 ms \citep{Metzger17}. 

\subsection{Hypernebula} \label{subsec:hypernebula}

Another paradigm proposed to explain PRSs associated with FRBs is the `hypernebula' model, which involves a stellar-mass NS/BH accreting at super-Eddington rates from a stellar companion, leading to intense mass-loss which inflates a nebula of relativistic electrons \citep{Sridhar+22, Sridhar+24a}. This model has been used to self-consistently explain the observed properties of the PRSs associated with FRBs\,20121102A, 20190520B \citep{Bhandari+23b}, as well as the radio bursts themselves \citep{Sridhar+21}. This model is also consistent with the burst properties for FRBs\,20210117A \citep{Bhandari+23a} and 20201124A \citep{Dong24}, and the limits on PRS emission at their locations. Additionally, the proposed hypernebula resembles the Galactic microquasar SS 433 -- W50 system \citep{Dubner98}.

Applying the model to source J1136$+$2643, we consider a BH of mass $M_\bullet=10\,M_\odot$ accreting matter from a companion donor star of mass M$_\star$ = $30\,M_\odot$ at a super-Eddington rate $\dot{M}$. Such mass transfer is expected to drive large-angled slow winds from the accretion disk, which interact with the upstream circumstellar medium (with an assumed density $n_{\rm csm}\sim 10\,{\rm cm^{-3}}$) and accumulate an ejecta shell. The kinetic luminosity of the disk-wind with velocity $v_{\rm w}$ is $L_{\rm w}$ $\approx\frac{1}{2}\dot{M}v_{\rm w}^2$ (Equation~7 in \citealt{Sridhar+22}). In addition to the slower disk winds, the accreting BH launches a faster, collimated jet along the spin axis (with speeds $v_{\rm j}\gg v_{\rm w}$), corresponding to a jet luminosity $L_{\rm j}\approx\eta L_{\rm w}$, with the jet efficiency being $\eta<1$. The radio synchrotron emission in the hypernebula arises from the collision between the collimated jet and the slower disk-wind shell that heats relativistically hot electrons behind the shell. The luminosity of the shock-heated electrons can be taken as $L_{\rm e}\approx\varepsilon_{\rm e} L_{\rm j}$, where $\varepsilon_{\rm e}\sim0.5$ is the heating efficiency of the electrons \citep{Sironi&Spitkovsky_11} in a mildly-magnetized electron-ion shock with jet magnetization $\sigma_{\rm j}\sim 0.1$.

\begin{figure}
    \centering
    \includegraphics[width=0.48\textwidth]{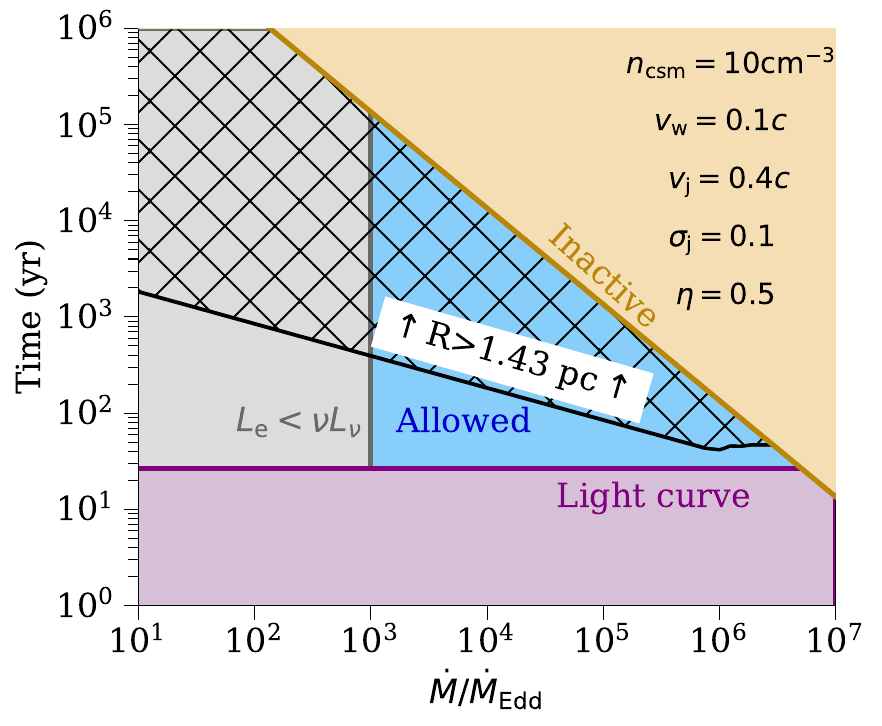} 
    \caption{Hypernebula model constraints in accretion rate-evolutionary age parameter space for the Tier A source J1136$+$2643 where the accretion rates are normalized to the Eddington accretion rate $\dot{M}_{\rm Edd}\equiv L_{\rm Edd}/(0.1c^2)$, where $L_{\rm Edd}\simeq1.5\times10^{39}\,{\rm erg\,s^{-1}}(M_\bullet/10\,M_\star)$ is the Eddington luminosity. The assumed fiducial set of parameters for the system is listed in the top right of the figure. The blue shaded area is the only region that likely represents the putative hypernebula associated with J1136$+$2643 based on the observed constraints on the size, luminosity, radio SED, and light curve (more on this in \S\ref{subsec:hypernebula}).}
    \label{fig:hypernebulae}
\end{figure}

We place constraints on the accretion rate $\dot{M}$ (normalized to the Eddington accretion rate for a $10\,M_\odot$ BH) and evolutionary timescale of a putative hypernebula using the size constraint and luminosity of source J1136$+$2643 in Figure~\ref{fig:hypernebulae}. We adopt a fiducial set of parameters as listed in the top right of the panel. First, accretion onto the BH by the donor star sets an active lifetime for the hypernebula (brown shaded region) given by Equation~3 of \cite{Sridhar+22}. Next, using the available constraint on the physical size of J1136$+$2643 ($<1.43$\,pc) and Equation~10 of \cite{Sridhar+22} which bridges the free-expansion phase and deceleration phases of the ejecta, we rule out the black hatched region. We can further eliminate the parameter space corresponding to $t\lesssim$ 27 years based on the 1.4\,GHz light curve  (purple region). This eliminated region also includes the values of $\dot{M}$ and the ages of the system that yield inefficient synchrotron emission (i.e., a synchrotron cooling timescale $>$ the adiabatic cooling timescale; Equation~49 of \citealt{Sridhar+22}) and optical depth of free-free absorption ($\tau_{\mathrm{ff}}$) $<1$ at $\nu_b$ (3.6 GHz). Finally, using the observed radio luminosity of J1136$+$2643, we rule out the $\dot{M}$ range where $L_{\rm e}\lesssim2\times10^{38}\,{\rm erg\,s^{-1}}$ (grey region).

The remaining narrow blue region in Figure~\ref{fig:hypernebulae} denotes the allowed parameter space for Tier A source J1136$+$2643. For low accretion rates of $\dot{M}/ \dot{M}_{\rm Edd}$ $\lesssim 10^{4}$, the hypernebula can be considerably older $\sim$ 30 -- 300 years. Given the compact nature of the source, however, it instead suggests that the nebula is likely younger with a moderate mass-transfer rate. While a high accretion rate of $\dot{M}/ \dot{M}_{\rm Edd} \approx 10^{5-6}$ is possible, the nebula age is confined to a very narrow range of $\sim$ 30 -- 50 years.

\subsection{Energy Equipartition}\label{subsec:equipartition}

Independent of any progenitor models, we can use the SSA feature in the radio SED of J1136$+$2643 to place meaningful constraints on the source properties. Specifically, for self-absorbed emission, we can indirectly constrain the size $R$, magnetic field $B$, and magnetic energy $U$ of the emitting region under the assumption of energy equipartition between radiating particles and the magnetic field \citep{Scott77, Chevalier98}.

For an observed flux density of 2 mJy at $\nu_{b}$, we estimate a source size of $R= 0.05$~pc, consistent with the upper limit of 1.43 pc inferred from VLBA observations \citep{Sargent22}. In a similar fashion, we estimate a magnetic field strength of $B$ $=$ 0.28~G (Equation~14 in \citealt{Chevalier98}) and a corresponding magnetic energy of $U \sim 10^{49}~\rm erg$. For those calculations, we assume a fiducial value for the emission volume filling factor of $f = 0.5$, a ratio of relativistic electron energy density to magnetic energy density of $\alpha = 1$, and an energy conversion efficiency parameter of $\epsilon_{B}$ = $\frac{1}{3}$. 

The inferred magnetic strength is a factor of $\sim10^2$ higher than the average magnetic field strengths of PRSs\,20121102A and 20190520B derived from the RM and host contribution to DM ($\approx$ 1 -- 6~mG; \citealt{Michilli18, Anna-Thomas23}). For reference, the host contribution of the integrated magnetic field strength along the line of sight for a sample of 10 non-repeating FRBs is on the order of $\mu$G \citep{Sherman23}. We stress that these estimated values for the known FRB-PRS limits are only lower limits as the host DM is likely dominated by star formation within the host rather than the immediate surrounding environments. For comparison, our inferred magnetic field strength is similar to that expected from the magnetar wind nebulae model of PRS 20121102A (e.g., Equation 17 in \citealt{Margalit18}). The inferred energy, implies an average energy-injection rate $\dot{E} \gtrsim 10^{40} \, {\rm erg \, s}^{-1}$ over the $> 30\,{\rm yr}$ age of the source.

\section{Alternative origin for the radio emission: On a possible IMBH origin} \label{sec:AGNorigin}

Based on our findings presented in Sections \ref{sec:res} and \ref{sec:models}, we have identified source J1136$+$2643 as our most compelling FRB-PRS candidate. However, we cannot rule out the possibility that source J1136$+$2643 and the remaining sources in our sample have an alternative origin. In particular, as mentioned in Section \ref{sec:sample}, our sample of radio sources were first suggested to be ``wandering'' accreting IMBHs by \cite{Reines20} based on their compact nature and apparent offset from their host centers. This motivates us to consider their origin in the context of IMBHs as there are also several notable features in the observed SEDs of the radio sources which may corroborate such an interpretation.

For the Tier A source, J1136$+$2643, the SSA turnover feature shown in the SED (Figure \ref{fig:SED}), is similar to the spectral breaks observed in gigahertz peaked spectrum (GPS) and compact steep-spectrum sources, which exhibit a defining break frequency between 1 -- 5\,GHz from young AGN \citep{ODea98, Murgia03}. The observed turnovers in the radio SEDs of GPS sources are generally ascribed to SSA \citep{Snellen00}, driven by the compact size of AGN cores. The presence of such a feature in the radio SED of J1136$+$2643, also from SSA, combined with the compact size, are thus broadly consistent with a GPS interpretation with an AGN origin.

We now turn to Tier B sources, which primarily exhibit flat spectral indices ($\alpha \approx -0.20~{\rm to}~-0.03$), reminiscent of the flat spectrum radio cores of AGN. Indeed, it is widely accepted that the radio emission from AGN is anisotropic and orientation-dependent as evidenced by the discoveries of relativistic jets launched by the central BHs \citep{Urry95}. In this framework, the compact base of the approaching jet observed close to the jet-axis and the BH-accretion disk nucleus as viewed down the jet axis both result in a flat SED \citep{DeZotti2010}. We note that the Tier B source SEDs are similar to that of the known compact AGN source J1126$+$1252 as well as the likely AGN source J1027$+$1027 (also compact) in our sample. Thus, the spectral similarities, together with the large host-normalized offsets, suggest a potential association between Tier B sources and background AGN. It is worth noting that the compact sizes and large offsets in Tier B sources are also in line with the ``wandering'' IMBH scenario in which the radio source exhibits a large offset from the dwarf galaxy center. However, as discussed in \cite{Reines20}, it is difficult to distinguish between background interlopers and ``wandering'' IMBHs based only on radio observables.

If, however, the AGN is viewed in the plane of the disk, then the radio emission would instead be dominated by the radio lobes, which exhibit a steep radio spectrum \citep{DeZotti2010}. This is analogous to the observed steep spectral indices in our Tier C sources. Moreover, the non-compact nature of these sources is consistent with the extended jet structure in this scenario. This interpretation is further reinforced by their similarity to the known AGN source J1220$+$3020 in our sample, which exhibits a steep spectral index and extended radio emission.

Thus, while we identify a number of similarities between the radio sources in our sample and known FRB-PRSs, as well as viable models for their emission, we cannot discount their resemblance to ``wandering'' BHs and background AGN, which can also exhibit a diversity of radio SEDs. Hence, it is imperative to obtain follow-up observations in other wavelengths such as X-ray and optical bands, as they offer additional BH diagnostics when searching for PRSs coincident with FRBs. This is highlighted by the identification of at least two known AGN in our sample through optical observations. Similarly, the discovery of FRBs at the locations of these radio sources would provide unambiguous evidence for their origins and association with FRBs. 

\section{Conclusions}\label{sec:conclusions}

We have presented VLA and EVN observations of \nsamp~FRB-PRS candidates in dwarf galaxies at low redshifts ($z \lesssim 0.055$), two radio sources of known AGN origin, and one radio source with a probable AGN origin. All sources that were observed are detected with the VLA and EVN (where applicable).  We divided the sample into three tiers based on their observed properties and shared similarities with known FRB-PRS pairs, and further assessed the likelihood that any of the sources share a common origin with PRSs associated with FRBs\,20121102A and 20190520B. We also tested the viability of NS wind nebula and hypernebula models for our most promising FRB-PRS candidate J1136$+$2643. Our main conclusions are as follows:

\begin{itemize}
    \item We identify a single source (J1136$+$2643) as the most promising FRB-PRS candidate in our sample, classified in Tier A (compact on milliarcsecond scales and small galactocentric offset). Additionally, we identify two sources (J0019$+$1507 and J0909$+$5655) in Tier B (compact) and six sources (J0106$+$0046, J0903$+$4824, J0931$+$5633, J1200$-$0341, J1226$+$0815, and J1253$-$0312) in Tier C (not obviously compact based on existing observations). Source J0019$+$1507 is detected, but unresolved, in our EVN observations. 

    \item The radio sources exhibit host-normalized offsets of 0~r$_{e}$ $\leqslant$ r$_\mathrm{norm}$ $\leqslant$ 5.3~r$_{e}$. The Tier A source has r$_{\rm norm}=1.31 ~\rm r_{e}$, consistent with those of PRSs\,20121102A and 20190520B, whereas both Tier B sources have r$_\mathrm{norm}$ $> 3~r_{e}$, which may indicate a background AGN origin. While not compact, almost all Tier C sources are within 2~r$_{e}$ of the dwarf hosts, comparable to the host-normalized offsets of FRB-PRSs and transients J1419$+$3940 and VT1137$-$0337.

    \item The peak luminosity at $\nu_{b}$ of the Tier A source, J1136$+$2643, is consistent with NS wind nebula models. Within the magnetar nebula model we constrain the energy injection rate and ambient density. From this, we are able to rule out the AIC scenario as a potential formation pathway, while a core-collapse SNe or BNS merger origin remain plausible. In the hypernebula paradigm, constraints derived from the compact size ($<$ 1.43~pc) and spectral luminosity of J1136$+$2643 imply a nebula age of $\sim$ 50 yr for moderate $\dot{M}/\dot{M}_{\rm Edd}$, slightly older than the inferred ages of FRB-PRSs. Despite some differences from the known FRB-PRSs, the observed properties of J1136$+$2643 are consistent with multiple FRB progenitor models.

    \item The spectral radio luminosities of the sample span roughly there orders of magnitude ranging between $ \approx 10^{27}$ and 10$^{30}$ erg s$^{-1}$Hz$^{-1}$. Across all tiers, the radio sources are sub-luminous relative to the known FRB-PRSs (with the exception of source J0909$+$5633 in Tier C). Moreover, the luminosities are comparable to those of VT 1137$-$0337 (PWN candidate) and PTF10ghi (SLSN), but brighter than the decades-long transient J1419$+$3940. Additionally, the radio luminosities for our sample are comparable to existing PRS limits for well-localized FRBs, suggesting that these FRBs could harbor lower luminosity PRSs.

    \item On a temporal baseline of $\approx 30$ years, the light curves at 1.4\,GHz of all radio sources are generally flat or gradually decaying, with the Tier A source exhibiting slight variability. At 3 and 9\,GHz, the short-term light curves of the full sample remain relatively constant with no evidence of variability, comparable to the observed behaviors in the light curves of the known FRB-PRSs. Given the sparse and uneven data coverage across all frequencies, the diversity in the light curve behavior, and the inhomogeneous comparison timescales, the light curves do not offer obvious distinguishing power in determining the origin(s) of the radio sources in our sample.

    \item The turnover observed in the radio SED of Tier A source, J1136$+$2643, at $\nu_{b} = $ 3.6\,GHz can likely be attributed to SSA. This spectral shape contrasts with what is observed in the SED of PRS\,20121102A at $\sim$ 9\,GHz, caused by synchrotron cooling, but aligns with turnovers seen in GPS sources. In Tier B, the presence of flat spectral indices ($\alpha \approx -0.1$) is broadly consistent with flat-spectrum PWNe, as well as core-dominated radio AGN. Conversely, the steep spectral indices ($\alpha$ $\approx -0.5~{\rm to}~-1.0$) observed for the Tier C sources are reminiscent of extended radio lobes associated with AGN.
    
\end{itemize}

With only a couple of definitive FRB-PRSs to date, both identified through the initial localization of an FRB, it is important to explore alternative avenues for new identifications of FRB-PRSs to fully delineate the characteristics of the population and their connection to FRB progenitors. Through multi-frequency radio observations, we find that even our most promising source, J1136$+$2643, differs from known FRB-PRSs in terms of spectral evolution and radio luminosity, potentially indicative of diversity in the FRB-PRS population. Moreover, while we identify a number of similarities in the observed properties between our sample and known FRB-PRSs, we do not discount the fact that they could also be consistent with wandering BHs and background AGN. 

Our work has highlighted that even among a sample with a fairly uniform selection of compact radio sources in spectroscopically confirmed dwarf galaxies, there is great diversity in observed radio spectral and temporal behavior that can be challenging to interpret. In addition, this picture can be further complicated if the sample stems from multiple source populations. Nevertheless, there remains great flexibility in current FRB-PRS emission models and mechanisms. This highlights the importance of high-resolution VLBI observations to confirm the compactness of candidate radio sources, coupled with X-ray observations and optical spectroscopy, to complement future radio searches of FRB-PRSs and reduce contamination from background interlopers. At the same time, the increased attention to blind searches of radio sources on many timescales from ThunderKAT, VLASS, and FRATS \citep{Fender16, TerVeen19, Law23} will place these compact radio sources into context with the entire source population.

\section{Acknowledgments}
The authors are grateful for the helpful discussions with Dillon Dong and Joseph Michail. Y.D. is supported by the National Science Foundation Graduate Research Fellowship under grant No. DGE-2234667. The Fong Group at Northwestern acknowledges support by the National Science Foundation under grant Nos. AST-2206494, AST-2308182, and CAREER grant No. AST-2047919. T.E. is supported by NASA through the NASA Hubble Fellowship grant HST-HF2-51504.001-A awarded by the Space Telescope Science Institute, which is operated by the Association of Universities for Research in Astronomy, Inc., for NASA, under contract NAS5-26555. S.B. is supported by the Dutch Research Council (NWO) Veni Fellowship (VI.Veni.212.058). Research by the AstroFlash group at the University of Amsterdam, ASTRON, and JIVE is supported in part by a Vici grant from the Dutch Scientific Research Council (Nederlandse Oranisatie voor Wetenschappelijk Onderzoek or NWO; PI Hessels; VI.C.192.045) as well as the European Union's Horizon 2020 research and innovation programme through the ERC Advanced Grant `EuroFlash' (PI Hessels; grant Agreement No. 101098079). W.F. gratefully acknowledges support by the David and Lucile Packard Foundation, the Alfred P. Sloan Foundation, and the Research Corporation for Science Advancement through Cottrell Scholar Award \#28284. N.S. acknowledges the support from NASA (grant number 80NSSC22K0332), NASA FINESST (grant number 80NSSC22K1597), a Columbia University Dean's fellowship, and a grant from the Simons Foundation. A.E.R. acknowledges support provided by the NSF through CAREER award 2235277 and NASA through EPSCoR grant No. 80NSSC20M0231. B.M. acknowledges financial support from the State Agency for Research of the Spanish Ministry of Science and Innovation under grant PID2019-105510GB-C31/AEI/10.13039/501100011033 and through the Unit of Excellence Mar\'ia de Maeztu 2020--2023 award to the Institute of Cosmos Sciences (CEX2019- 000918-M). The Berger Time-Domain Group at Harvard is supported by NSF and NASA grants.

%VLA
The National Radio Astronomy Observatory is a facility of the National Science Foundation operated under cooperative agreement by Associated Universities, Inc.
%EVN
The European VLBI Network is a joint facility of independent European, African, Asian, and North American radio astronomy institutes. Scientific results from data presented in this publication are derived from the following EVN project code(s): EO018.
%MMT and Keck
W. M. Keck Observatory and MMT Observatory access was supported by Northwestern University and the Center for Interdisciplinary Exploration and Research in Astrophysics (CIERA). The authors wish to recognize and acknowledge the very significant cultural role and reverence that the summit of Maunakea has always had within the indigenous Hawaiian community. We are most fortunate to have the opportunity to conduct observations from this mountain.
\vspace{5mm}
\software{\texttt{AIPS} \citep{AIPS}, \texttt{Astropy} \citep{Astropy13, Astropy18, Astropy22}, \texttt{CASA} \citep{casa, CASA22}, \texttt{DIFMAP} \citep{DIFMAP}, \texttt{GALFIT} \citep{galfit_2002, galfit_2010}, \texttt{Heimdall}, \texttt{Matplotlib} \citep{Hunter07}, Numpy \citep{Harris20}, \texttt{PHOTUTILS} \citep{photutils}, \texttt{SAOImage DS9} \citep{ds9}, \texttt{SciPy} \citep{2020SciPy-NMeth}}, \texttt{SFXC} \citep{Keimpema+15} 

\facilities{VLA, EVN, Keck (DEIMOS), MMT (Binospec)}

\begin{figure}[h]
\centering
\hspace*{-0.25cm}
\includegraphics[width=0.8\textwidth,trim={5cm 5cm 5cm 5cm}]{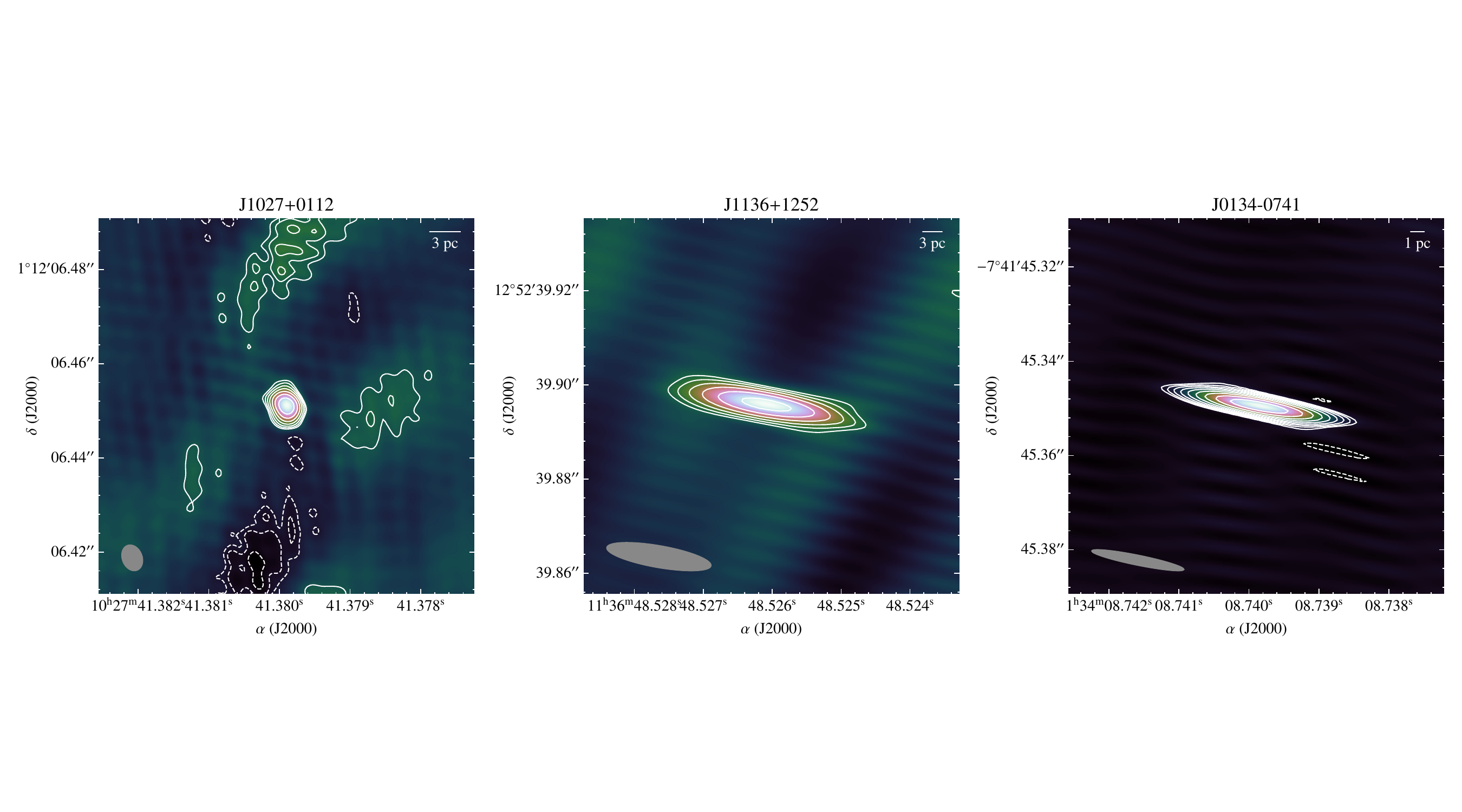}
\caption{EVN images of the identified likely and confirmed background AGN sources, J1027$+$0112 and J1136$+$1252, along with the additional source J0134$-$0741 at 1.7\,GHz with contour lines depicting rms levels starting from $3.5\sigma$. The synthesized beams are displayed in the bottom left corner in each panel with sizes {6\,mas $\times$ 4\,mas}, {22\,mas $\times$ 5\,mas}, {24\,mas $\times$ 3\,mas} for J1027$+$0112, J1136$+$1252, and J0134$-$0741, respectively.}
\vspace{-5pt}
\label{fig:extra_EVN}
\end{figure}

\bibliography{refs}
\appendix \label{sec:appendix}
We also observed the source J0134$-$0741 as a part of our EVN observations. This source is included in sample B of the \citet{Reines20} sample of compact radio sources and therefore does not satisfy our selection criteria for further analysis. The source is detected with the EVN with a flux density of $10.0\pm1.5$\,mJy at 1.7\,GHz. We constrain the angular size of the source to be $< 1.9$\,mas which translates to a physical size of $<0.62$\,pc at $z=0.0156$. We measure the luminosity of the source to be $5.8 \times 10^{28}$ erg\,s$^{-1}$\,Hz$^{-1}$. We show the EVN radio image of source J0134$-$0741 and the identified probable and confirmed background AGN sources J1027$+$0112 and J1136$+$1252 in Figure\,\ref{fig:extra_EVN}.

\end{CJK*}
\end{document}